\documentclass{aastex631}
\usepackage{amsmath,amssymb,lipsum}
\usepackage{framed}
\usepackage{natbib}

\begin{document}

	\setcounter{page}{1}
	
	\pagestyle{plain}

	\setcounter{page}{1}
	
	\pagestyle{plain}
	
	\title{Cosmological Implications of the Extended Uncertainty Principle: Energy Conditions, Stability, and Late Time Acceleration}
	\author{M. Roushan}\email{m.roushan@umz.ac.ir(Corresponding Author)},\author{N. Rashidi}\email{n.rashidi@umz.ac.ir}
	\affiliation{Department of Theoretical Physics, Faculty of Science,
		University of Mazandaran,\\
		P. O. Box 47416-95447, Babolsar, IRAN}

	\begin{abstract}
		We study the cosmological consequences of the Extended Uncertainty Principle (EUP) by deriving modified Friedmann equations through thermodynamic arguments. The evolution of the effective equation of state induced by EUP corrections is analyzed and characterized using the Chevallier–Polarski–Linder (CPL) parametrization. We then examine the fulfillment of classical energy conditions, including the null, weak, strong, and dominant conditions. The dynamical and thermodynamic stability of the model is investigated, showing that the EUP cosmology admits a late-time de Sitter attractor. Finally, we evaluate the effective speed of sound associated with the model and discuss implications for perturbative stability. Our findings indicate that EUP-induced corrections can produce a consistent late time acceleration without requiring a cosmological constant.
	\end{abstract}
	
	\section{Introduction}
	
	A wealth of observational evidence including measurements of Type Ia supernovae, cosmic microwave background (CMB) anisotropies, and baryon acoustic oscillations strongly indicates that the universe is currently undergoing a phase of accelerated expansion \cite{Planck2018,Hicken2009,Weinberg}. While the cosmological constant \( \Lambda \) provides a straightforward explanation within the framework of the \(\Lambda\)CDM model, its associated fine-tuning and coincidence problems have motivated a wide spectrum of theoretical investigations into dynamical dark energy and modified gravity models \cite{Sahni2008,Capozziello2020,Wald2001}.
	
	One promising direction involves the thermodynamic interpretation of gravity, in which spacetime dynamics emerge from horizon thermodynamics. Pioneering work by Jacobson demonstrated that Einstein’s field equations can be derived from the Clausius relation, linking heat flow to entropy change across local Rindler horizons \cite{Jacobson1995}. This perspective was extended by Padmanabhan and Verlinde, who argued that gravity may be an emergent phenomenon rooted in statistical and thermodynamic principles \cite{Padmanabhan2005,Verlinde2011}. Central to this view is the black hole entropy–area law, originally formulated by Bekenstein and Hawking \cite{Bekenstein1973,Hawking1975}, which provides a quantum gravitational modification to classical notions of geometry and energy.
	
	Recent studies have explored how quantum gravitational effects particularly those modifying the Heisenberg uncertainty principle can influence large scale cosmological dynamics. The \textit{Generalized Uncertainty Principle (GUP)} introduces a minimal measurable length scale and has been extensively explored in high energy, black hole, and cosmological contexts 
	\cite{Amati1989,Maggiore1993,Garay1995,Kempf1995,Scardigli1999,Adler2001,Das2008,Ali2009,Hossenfelder2013,Roushan2014,Tawfik2015,Carr2015,Carr2016,
		Roushan2016,Roushan2018,Casadio2020,Carr2022,Bosso2023}. 
	Conversely, its infrared counterpart, the \textit{Extended Uncertainty Principle (EUP)}, introduces a minimal measurable momentum, leading to an infrared cutoff that modifies spacetime thermodynamics and large scale cosmological evolution 
	\cite{Hinrichsen1996,Mignemi2010,Dabrowski2019,Mureika2019,Chung2019,Roushan2019,Roushan2020,Lambiase2021,Pantig2022,Rashidi2023,Pantig2025}.
	
	By incorporating EUP-inspired corrections into the entropy–area relation at the apparent horizon of a Friedmann–Robertson–Walker (FRW) universe, it is possible to derive modified Friedmann equations containing additional terms that arise from this infrared deformation. In previous work, it was shown that a negative EUP deformation parameter \( \eta < 0 \) is sufficient to induce late-time acceleration without invoking dark energy or exotic scalar fields \cite{Roushan2024}.
	
	In this study, we extend this framework to examine the deeper physical viability of EUP-modified cosmology. We analyze the model’s compatibility with classical energy conditions (null, weak, strong, and dominant), investigate its dynamical and thermodynamic stability, and assess the perturbative behavior via the effective sound speed. A key aspect of our analysis involves characterizing the redshift evolution of the effective equation-of-state parameter \( w(z) \), which governs the relation between pressure and energy density. To facilitate comparison with observational constraints, we project this evolution onto the Chevallier–Polarski–Linder (CPL) \( (w_0, w_a) \) parameterization \cite{Chevallier2001,Linder2003}, where \( w_0 \) is the present-day value of the EoS and \( w_a \) encodes its evolution with scale factor. This widely used approach allows our quantum gravity-motivated model to be benchmarked against empirical bounds from Planck, BAO, and supernova data.
	
	Our findings reveal that the EUP corrections naturally drive a transition from a decelerated to an accelerated expansion phase and allow for a smooth crossing of the phantom divide \( w = -1 \). For values of the deformation parameter \( \eta \sim \mathcal{O}(10^{-27}) \), the model yields \( (w_0, w_a) \) pairs that lie within current observational confidence intervals, demonstrating its cosmological viability. Moreover, the fluid derived from EUP-induced corrections satisfies the null, weak, and dominant energy conditions, while violating the strong energy condition an expected trait of accelerating universes. The model also exhibits stable phase-space behavior and a positive, finite sound speed, confirming its dynamical and thermodynamic robustness.
	
	This paper is organized as follows: In Section~2, we introduce the EUP framework and derive the modified Friedmann equations. Section~3 examines the behavior of the effective EoS parameter and its projection onto the \( (w_0, w_a) \) plane. In Section~4, we assess the model’s compatibility with classical energy conditions. Section~5 analyzes the dynamical and thermodynamic stability of the cosmology. Finally, Section~6 summarizes our results and discusses their broader implications.
	
	\section{Extended Uncertainty Principle (EUP) Model}
	The standard Heisenberg Uncertainty Principle (HUP), represents a cornerstone of quantum mechanics. However, the incorporation of gravity into quantum frameworks has long suggested the necessity of modifying this relation. While high energy (ultraviolet) modifications have been extensively explored through the Generalized Uncertainty Principle (GUP), the large scale (infrared) regime remains comparatively underdeveloped. One natural extension in this direction is the Extended Uncertainty Principle (EUP), which introduces a long wavelength cutoff and has received increasing attention in the context of infrared modified gravity, black hole thermodynamics, and cosmology.
	In the EUP framework, a fundamental minimal measurable momentum arises as an infrared (IR) counterpart to the minimal length in GUP scenarios. This concept can be formalized through a deformation of the canonical position-momentum commutation relation. In the present analysis, we adopt a simplified isotropic realization of the EUP algebra~\cite{Hinrichsen1996,Mignemi2010}
	\begin{equation}\label{eq1}
		[X_i,P_j]=i\hbar(\delta_{ij}+\eta_{ijkl}x^kx^l+...)\,,
	\end{equation}
	where $\eta$ is a small deformation parameter. This deformation implies that the effective position and momentum operator takes the form
	\begin{eqnarray}\label{eq2}
		X_i=x_{i}\,,\quad P_i=p_i(1+\eta x^2)\,,
	\end{eqnarray}
	with $[x_i,p_j]=i\hbar\delta_{ij}$ defining the canonical structure. The parameter $\eta$, which has units of inverse length squared, introduces an IR modification that becomes relevant at cosmological scales.
	From Eq. (1), one obtains the corresponding uncertainty relation
	\begin{equation}\label{eq3}
		\Delta X \Delta P\geq\frac{\hbar}{2}(1+\eta (\Delta x)^2)\,,
	\end{equation}
	which, unlike the standard HUP, is not minimized when $\Delta x\rightarrow 0$, but instead admits a nonzero lower bound on momentum uncertainty. By minimizing the right hand side of Eq. (3), we obtain the minimal measurable momentum
	\begin{equation}\label{eq4}
		\Delta p_{min}=\hbar\sqrt{|\eta|} \,.
	\end{equation}
	This condition indicates the existence of an infrared cutoff (a finite lower limit on momentum fluctuations) with significant implications in gravitational contexts. In particular, a negative $\eta$ yields a physically consistent framework: a universe characterized by a minimum momentum scale that remains fully compatible with quantum mechanics. This insight underpins the cosmological model developed in this work.
	
	Figure 1 depicts the modified uncertainty relation from Eq.~(\ref{eq3}), showing $\Delta p$ as a function of $\Delta x$ for various values of the deformation parameter $\eta$. The $\eta = 0$ curve reproduces the familiar hyperbola of the standard Heisenberg Uncertainty Principle (HUP), where $\Delta p$ diverges as $\Delta x \to 0$ and vanishes as $\Delta x \to \infty$. In contrast, for negative $\eta$, the uncertainty behavior changes qualitatively: $\Delta p$ no longer approaches zero at large $\Delta x$, but instead saturates at a finite minimal value, $\Delta p_{\text{min}}$, which increases with $|\eta|$. This saturation reflects the infrared (IR) modifications to momentum space introduced by the Extended Uncertainty Principle (EUP). For visual clarity, we use illustrative values of $\eta$ on the order of $10^{-1}$ in Figure 1 many orders of magnitude larger than the phenomenologically consistent value ($\eta \sim 10^{-27}$). This rescaling preserves the qualitative features of the curves while enhancing their visual distinction. The theoretical and observational results discussed later in the paper use physically realistic values of $\eta$. In essence, the EUP introduces a scale-dependent deformation of quantum uncertainty, marked by a nonzero minimal momentum. This IR modification becomes relevant at large length scales and offers a compelling theoretical basis for explaining late-time cosmic acceleration without resorting to exotic dark energy or modifications of general relativity. The following sections explore the resulting thermodynamic and cosmological implications in detail.
	
	\begin{figure}
		\centering
		\includegraphics[width=0.40\linewidth]{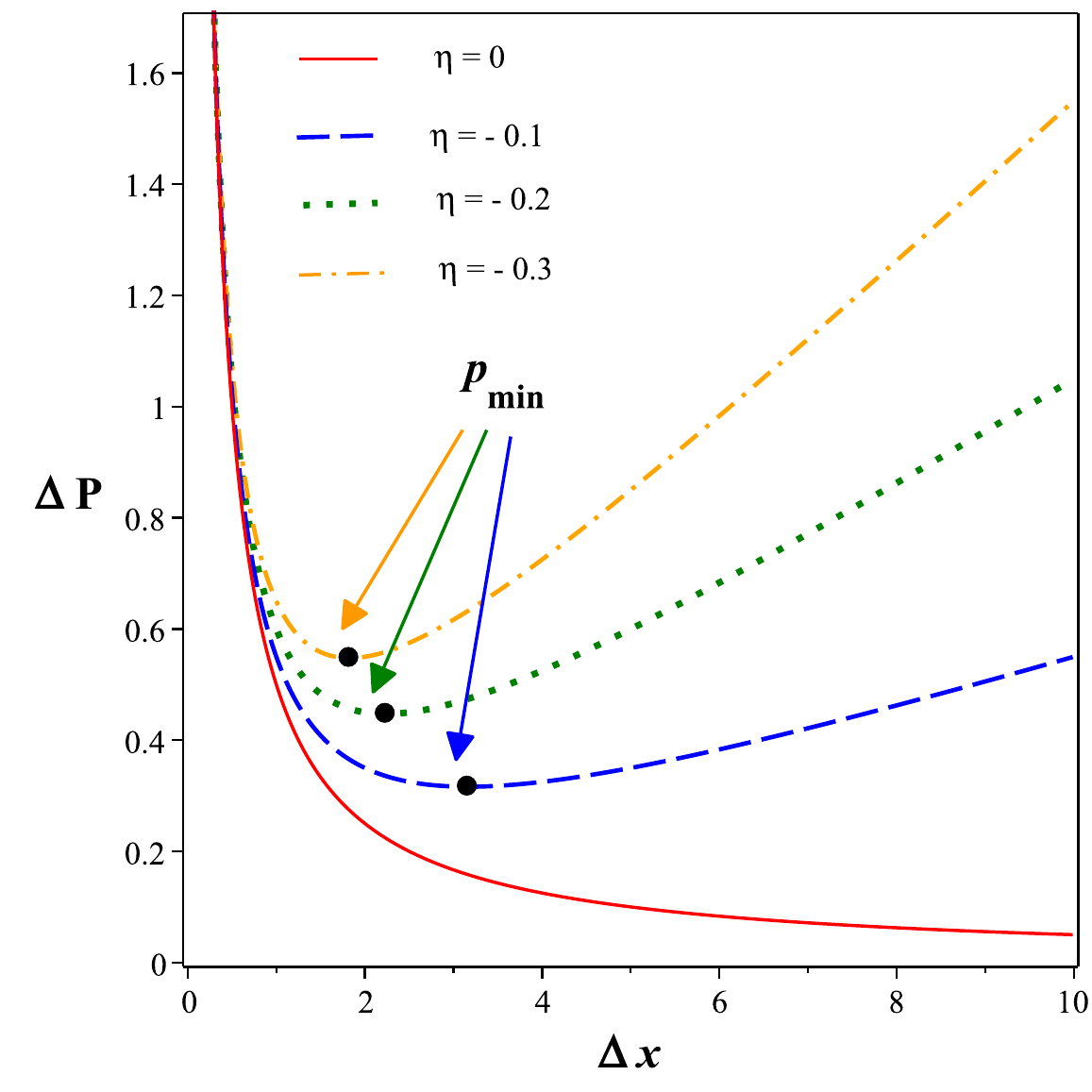}
		\caption{\small Standard and Extended Uncertainty Principle for different values of $\eta$ with specified minimal momentum points. For clarity of illustration, we have used larger values of $\eta$. The actual model employs $\eta\sim-10^{-27}$, consistent with the constraints used in our cosmological analysis in~\cite{Roushan2024}.}
		\label{fig1}
	\end{figure}
	
	In our previous work~\cite{Roushan2024}, we showed that incorporating the Extended Uncertainty Principle (EUP) through modifications to the entropy–area relation introduces significant corrections to the thermodynamic evolution of the apparent horizon in a spatially flat Friedmann–Robertson–Walker (FRW) universe. By applying the first law of thermodynamics to the apparent horizon,$dE=TdS+WdV$ (where $T$ is the horizon temperature (proportional to surface gravity), $S$ is the corrected entropy, and $W$ is the work density) we derived modified Friedmann equations that reflect these corrections. Notably, in the presence of a minimal measurable momentum (an infrared cutoff), the corrections become relevant at large scales, leading to an emergent dynamical effect that influences the universe’s late-time expansion.
	The modified Friedmann equations obtained from this thermodynamic treatment, including EUP corrections to the entropy, are given by
	\begin{eqnarray}
		\label{eq5}\frac{G}{2l_{pl}^2}H^2+\frac{G}{2\,l_{pl}^2}H\sqrt{H^2-16\eta}-\frac{8G\eta}{l_{pl}^2}
		\ln\Bigg(H+\sqrt{H^2-16\eta}\Bigg)
		=\frac{8\pi\,G}{3}\,\rho\,,
	\end{eqnarray}
	
	\begin{eqnarray}
		\label{eq6}\dot{H}\Bigg(\frac{8\,G\,\eta}{H^2\,l_{pl}^2}\Bigg)\Bigg(\frac{1}{1-\sqrt{1-\frac{16\eta}{H^2}}}\Bigg)=-4\pi\,G\,(\rho+p)\,.
	\end{eqnarray}
	Where, natural units ($\hbar=1=c$) are used throughout. These equations describe the cosmic dynamics in a universe influenced by an infrared deformation of the quantum uncertainty principle, consistent with the existence of a minimal measurable momentum.
	By identifying the additional terms in Eq. (\ref{eq5}) as an effective contribution to the energy budget, we define an EUP-induced energy density $\rho_{EUP}$  such that
	\begin{eqnarray}
		\label{eq7}H^2=\frac{8\pi\,G}{3}\Big(\rho+\rho_{EUP}\Big)\,,
	\end{eqnarray}
	yielding,
	\begin{eqnarray}
		\label{eq8}\rho_{_{EUP}}\equiv\frac{3}{8\pi\,G}\Bigg(\bigg(1-\frac{G}{2l_{pl}^2}\bigg)H^2-\frac{G}{2l_{pl}^2}H\sqrt{H^2
			-16\eta}+\frac{8\,G\eta}{l_{pl}^2}\ln\bigg(H+\sqrt{H^2-16\eta}\bigg)\Bigg)\,.
	\end{eqnarray}
	This term can be interpreted as a geometric contribution to the cosmic energy budget, arising purely from quantum gravitational modifications at large scales.\\
	These modifications to the Friedmann equations, derived from EUP-corrected entropy considerations, introduce an additional term in the energy budget, denoted as $\rho_{EUP}$ . This term may act effectively as a dynamical source influencing the expansion history of the universe, particularly at late times. While its structure resembles that of a dark energy component, it arises from the geometric and thermodynamic implications of the infrared deformation, without the explicit introduction of new fields. To better understand the physical role and observational viability of $\rho_{EUP}$, we proceed to examine its influence on key cosmological indicators beginning with the deceleration parameter $q(z)$, which directly encodes the transition from deceleration to acceleration in the universe's evolution.
	\subsection{Deceleration Parameter in the EUP-Corrected Cosmology}
	
	To quantify the acceleration of the universe, we consider the deceleration parameter $q$, which is defined as
	\begin{eqnarray}\label{eq9}
		q \equiv -1 - \frac{\dot{H}}{H^2},
	\end{eqnarray}
	where $ H = \dot{a}/a $ is the Hubble parameter and $ \dot{H} $  is its time derivative. A negative value of $q$ corresponds to accelerated cosmic expansion, while a positive value indicates deceleration.
	
	In the EUP framework with a negative deformation parameter $\eta<0$, the modification to the Friedmann equations leads to a nontrivial expression for $\dot{H}$, resulting from the correction to the entropy-area law at the apparent horizon. Solving the EUP-modified Friedmann equation thermodynamically, we obtain the following expression for the time derivative of the Hubble parameter
	\begin{eqnarray}
		\label{eq10}\dot{H}=-\frac{4\pi\rho_0\,l_{pl}^{2}\Biggl[{\cal{W}}\!\Big(-\frac{1}{16\eta}\exp\{{\frac{-2\pi\,l_{pl}^{2}\rho_0\left(1+z\right)^{3}
					+3\eta}{3\eta}}\}\Big)-1\Biggl]}{{\cal{W}}\!\left(-\frac{1}{16\eta}\exp\{{\frac{-2\pi\,l_{pl}^{2}\rho_0(1+z^3)+3\eta}{3\eta}}\}\right)}\left(1+z\right)^{3}\,,
	\end{eqnarray}
	
	where $\cal{W}$ denotes the \textit{Lambert function}, defined as the inverse function satisfying
	${\cal{W}}(x)e^{{\cal{W}}(x)} = x$. In this work, we use the principal real-valued branch ${\cal{W}}_0$,
	which ensures continuity and physical consistency for the argument domain encountered
	in our cosmological model. This function appears due to the transcendental nature of the entropy correction and encapsulates the implicit structure of the modified cosmic dynamics. Its presence ensures the correct analytic behavior of $\dot{H}$ across different cosmological epochs.
	
	Substituting Eq.~(\ref{eq12}) into Eq.~(\ref{eq11}), we obtain the EUP-corrected form of the deceleration parameter
	
	\begin{eqnarray}
		\label{eq11}q=\frac{\Biggl[-{\cal{W}}\!\Big(-\frac{1}{16\eta}\exp\{{\frac{-2\pi\,l_{pl}^{2}\rho_0\left(1+z\right)^{3}
					+3\eta}{3\eta}}\}\Big)+1\Biggl]\eta-\pi\,l_{pl}^{2}\rho_0\left(1+z\right)^{3}}{\Biggl[{\cal{W}}\!\Big(-\frac{1}{16\eta}\exp\{{\frac{-2\pi\,l_{pl}^{2}\rho_0\left(1+z\right)^{3}
					+3\eta}{3\eta}}\}\Big)-1\Biggl]\eta}\,.
	\end{eqnarray}
	This form reveals how the EUP deformation parameter $\eta$ affects the cosmic acceleration, with larger negative values of $\eta$ leading to stronger deviations from classical deceleration. The structure is regular and well-behaved for all $z\geq0$, ensuring physical consistency.
	
	\subsection{Effective Equation of State Parameter}
	
	The effective equation of state (EoS) parameter $w(z)$ serves as a diagnostic of the cosmic fluid’s nature and evolution. It is defined kinematically in terms of the deceleration parameter $q(z)$ via the relation
	\begin{eqnarray}
		\label{eq12}
		w = \frac{2}{3}q - 1.
	\end{eqnarray}
	which remains valid in the EUP framework provided $q(z)$ is derived from the modified Friedmann dynamics.
	
	In the present model, based on a negative deformation parameter $\eta < 0$, the infrared correction to the entropy-area relation at the apparent horizon introduces nontrivial modifications to the cosmic expansion. These corrections manifest through the Lambert $W$ function, and yield the following redshift-dependent form of the EoS parameter
	
	\begin{eqnarray}
		\label{eq13}w(z)=-\frac{2\pi\,l^{2}\rho_0\left(1+z\right)^{3}
		}{3\Bigg[{\cal{W}}\!\left(-\frac{1}{16\eta}\exp\{{\frac{-2\pi\,l_{pl}^{2}\rho_0\left(1+z\right)^{3}
					+3\eta}{3 \eta}}\}\right)-1\Bigg]\eta}-1
		\,.
	\end{eqnarray}
	
	This expression clearly encapsulates the quantum gravitational foundation underlying the cosmic dynamics. As illustrated in Figure~5 of our analysis in~\cite{Roushan2024}, the evolution of the equation-of-state parameter \( w(z) \) reveals a compelling transition: the universe shifts from a quintessence-like regime (\(-1 < w < -1/3\)) to a phantom phase (\(w < -1\)) at low redshift. Remarkably, this crossing of the phantom divide, \( w = -1 \), emerges naturally within the model—requiring neither exotic fields nor fine-tuning. This behavior aligns closely with observationally inferred trajectories, such as those reported by Planck 2018, which favor \( -1.06 < w < -1.00 \) at the 68\% confidence level.\\
	
	Such crossing behavior arises solely from the EUP deformation’s influence on standard matter and indicates a robust IR correction consistent with a late-time acceleration phase. For deformation parameters in the range $\eta \sim \mathcal{O}(10^{-27})$, the model’s predictions remain fully compatible with current datasets. Thus, the EUP-corrected equation of state supports the cosmological viability of the framework and provides a quantum gravity-based alternative to traditional dark energy paradigms.
	
	\subsection{Pressure and Energy Density Evolution}
	Having established the behavior of the effective equation of state parameter $w(z)$ and its remarkable transition across the phantom divide, we now turn to the explicit evolution of the underlying energy density and pressure components. This step is crucial for deepening our understanding of the physical content of the model and for assessing its compatibility with observational and theoretical constraints.
	Solving Eq. (\ref{eq5}) directly provides the total energy density of the universe as
	\begin{eqnarray}
		\label{eq14}\rho=\frac{3}{16\pi l_{pl}^2}\Bigg(H^{2}+H \sqrt{H^{2}-16 \eta}-16 \eta  \ln \! \left(H +\sqrt{H^{2}-16 \eta}\right)\Bigg)\,,
	\end{eqnarray}
	by using the energy density $\rho$ obtained in Eq. (\ref{eq14}) together with the effective equation of state parameter $w$ derived in Eq. (\ref{eq13}), and the standard relation between pressure and energy density $w=\frac{p}{\rho}$ we obtain an explicit expression for the pressure, given by
	
	\begin{eqnarray}
		\label{eq15}p = -\frac{ \left( 2\pi l_{pl}^{2} \rho_0 (1+z)^3
			+ 3\eta \left( \mathcal{W}\left( -\frac{1}{16\eta}
			\exp\left\{ \frac{-2\pi l_{pl}^{2} \rho_0 (1+z)^3 + 3\eta}{3\eta} \right\} \right)
			- 1 \right) \right) }{16\eta\pi l_{pl}^{2} \left(
			\mathcal{W}\left( -\frac{1}{16\eta}
			\exp\left\{ \frac{-2\pi l_{pl}^{2} \rho_0 (1+z)^3 + 3\eta}{3\eta} \right\} \right)
			- 1 \right) } \notag \\
		\times \left( H^2 + H\sqrt{H^2 - 16\eta}
		- 16\eta \ln\left( H + \sqrt{H^2 - 16\eta} \right) \right)
		\,.
	\end{eqnarray}
	
	\begin{figure}
		\centering
		\includegraphics[width=0.38\linewidth]{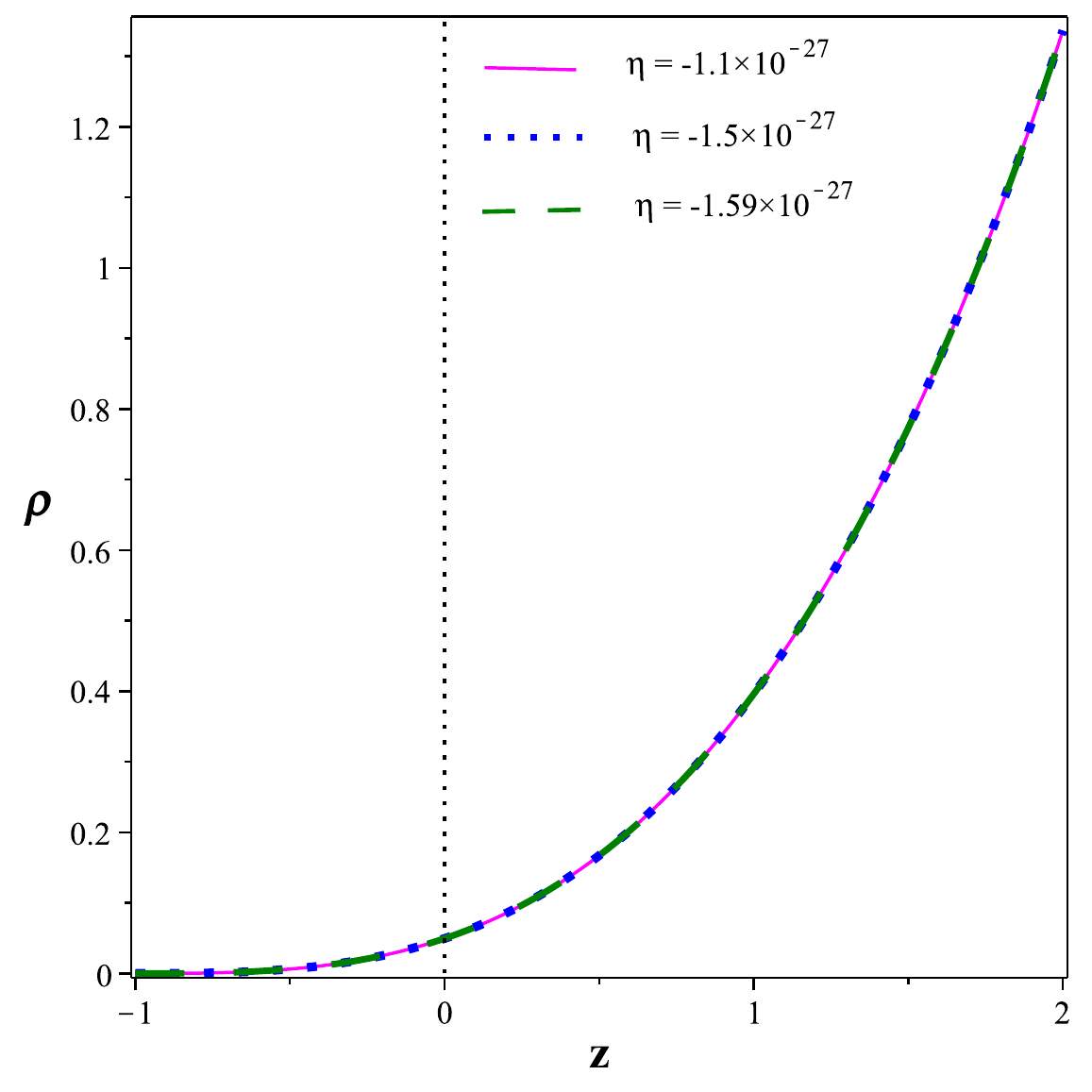}
		\includegraphics[width=0.38\linewidth]{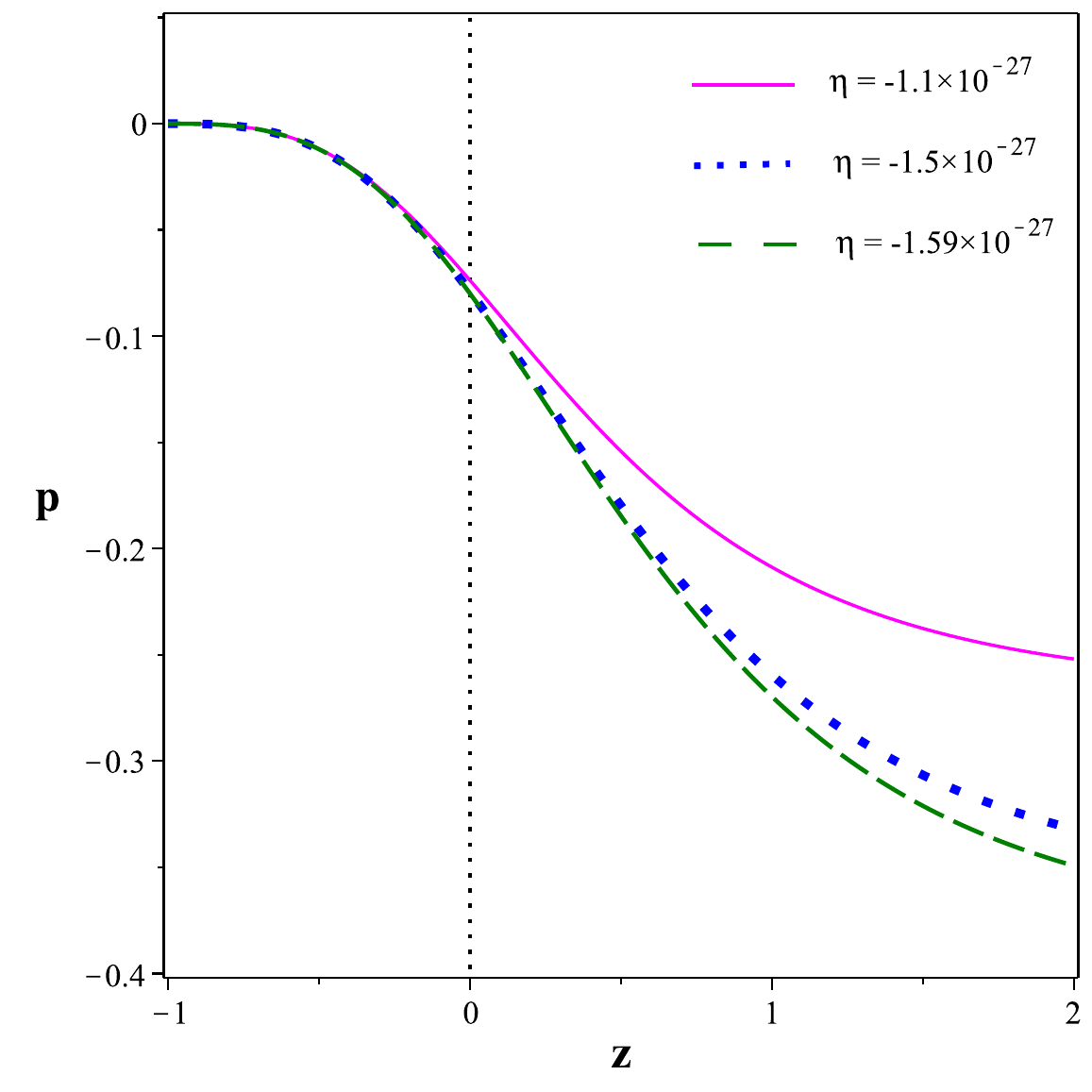}
		\caption{\small Energy density versus redshift (left panel) and pressure versus redshift
			(right panel) for different values of $\eta$ to have late-time cosmic speed-up..}
		\label{fig2}
	\end{figure}
	
	To further investigate the observational viability of this dynamical behavior, and to facilitate comparison with cosmological datasets, we proceed in the next section to map the EUP-derived $w(z)$ into the standard $(w_0,w_a)$ parameter space using the Chevallier–Polarski–Linder (CPL) formalism.
	
	To complement the redshift-based formulation of the equation of state, we also derive an exact expression for the EUP-induced equation of state parameter in terms of the scale factor a. By expressing the modified Friedmann equations in terms of $a=\frac{1}{1+z}$, and following the same thermodynamic procedure, we obtain
	\begin{eqnarray}
		\label{eq16}
		w(a)=-1-\frac{2 \pi  \,l_{pl}^{2} \rho_{_{0}}}{3 \left({\cal{W}}\! \left(-\frac{1}{16\eta}\exp\{-\frac{2 \pi  \,l_{pl}^{2} \rho_{_{0}}}{3\eta a^{3}}+1\}\right)-1\right) \eta  \,a^{3}}\,.
	\end{eqnarray}

	For a more detailed account of the derivation and related discussions, see~\cite{Roushan2024}.
	
	\section{Extended Uncertainty Principle Model in the $(w_0,w_a)$ Plane}
	We consider natural cutoffs as a generic dynamical effect in the context of EUP model, therefore its equation of state parameter $w\equiv \frac{p}{\rho}$  will be a function of time. In this respect, $p$ is the pressure and $\rho$ is the energy density of the energy density of matter, modified by quantum gravitational effects, drives the observed accelerated expansion of the universe. In these circumstances, the equation of state parameter is characterized as a function of the universe’s scale factor $a$, such that  $w = w(a)$. To investigate a time-based equation of state, we take up the functional structure as a generic Taylor expansion of $w$ in powers of the scale factor up to order $N$ as follows
	\begin{eqnarray}
		\label{eq17}w(a)=w_0+\sum_{i=1}^N(1-a)^{i}w_i\,.
	\end{eqnarray}
	Note that, all parameters are unfluctuating when adopting higher order polynomials. In fact, the $w_i$ parameters are limited and going from  $N = 1$ ( the linear element) to $N = 2$ (the quadratic element) to $N = 3$ (the cubic element) does not better the goodness of fit and its compatibility with $\Lambda$CDM is still maintained, which implies that a linear parameterization is enough in this situation. So, it is logical this series come to a stop at first order in $w(a)$.
	
	The Taylor expansion of $w$ at first order in the scale factor parameterized by
	\begin{eqnarray}
		\label{eq18}w(a)=w_0+(1-a)w_a\,.
	\end{eqnarray}
	To connect the EUP-based expression for $w(a)$ with the CPL parametrization introduced above, we extract the values of $w_0\equiv w(a=1)$ and $w_a\equiv-\frac{dw}{da}\mid_{a=1}$ directly from the analytic form of the equation of state derived in Eq.~(\ref{eq16}). These are given by
	
	\begin{eqnarray}
		\label{eq19}w_0=-1+\frac{-\frac{2 \pi  \,l^{2} \rho0}{3 \eta}+a -1}{a^{3} \left(W\! \left(-\frac{{\mathrm e}^{-\frac{2 \pi  \,l^{2} \rho0}{3 \eta  \,a^{3}}+1}}{16 \eta}\right)-1\right)}\,,
	\end{eqnarray}
	
	\begin{eqnarray}
		\label{eq20}w_a=\frac{1}{a^{3} \left(W\! \left(-\frac{{\mathrm e}^{-\frac{2 \pi  \,l^{2} \rho0}{3 \eta  \,a^{3}}+1}}{16 \eta}\right)-1\right)}\,.
	\end{eqnarray}
	
	These expressions represent the present-day value and the leading-order evolution of the EUP-induced equation of state. They depend explicitly on the deformation parameter $\eta$, and naturally reduce to $w_0=-1$, $w_a=0$ in the classical limit $\eta\rightarrow 0$. This allows for a direct comparison with observational constraints in the $(w_0,w_a)$ plane, as explored in the subsequent analysis.
	
	This configuration estimates the equation of state parameter in late time universe consistent with observable data, so that $a\sim 1$. Here, $w_0$ is the present value of the equation of state parameter and $w_a$ exhibits the time evolution of dark energy models. Practically, assumed selection of $w_0$ and $w_a$ will identify the time evolution of the density caused by the dark energy, $\rho_{DE}$. Through the Friedmann equations, the expansion rate of the universe can be specified as a function of time with this density. We can determine the values of the expansion rate at various eras, taking into account a collection of observations. Next, we identify the $w_0$ and $w_a$ and also their related uncertainties that best possible fit the observations.
	It is important to distinguish between the CPL-fitted values $(w_0,w_a)$, which are obtained by approximating $w(a)$ over an interval of redshift, and the instantaneous values $(\hat{w}_0,\hat{w}_a)$, which are computed directly from the exact EUP-based expression for $w(a)$ at $a=1$. While they often align closely in slowly evolving models, differences can emerge in scenarios with non-linear dynamics, as we explore in the following analysis.
	We consider here the case of a Taylor expansion  of $w$ at present time ($a=1$) with new parametrization as
	\begin{eqnarray}
		\label{eq21}w(a)=\hat{w}_0+(1-a)\hat{w}_a\,.
	\end{eqnarray}
	where, $\hat{w}_0\equiv w(a=1)$, $\hat{w}_a\equiv \frac{dw}{da}(a=1)$.
	In the following, we investigate the pairs of $(w_0,w_a)$, $(\hat{w}_0,\hat{w}_a)$, $(\hat{w}_0,w_0)$ and $(\hat{w}_a,w_a)$ of our model for some different values of $\eta$ parameter in the background of marginalized posterior distributions of the $(w_0,w_a)$.

	\begin{figure}
		\centering
		\includegraphics[width=0.40\linewidth]{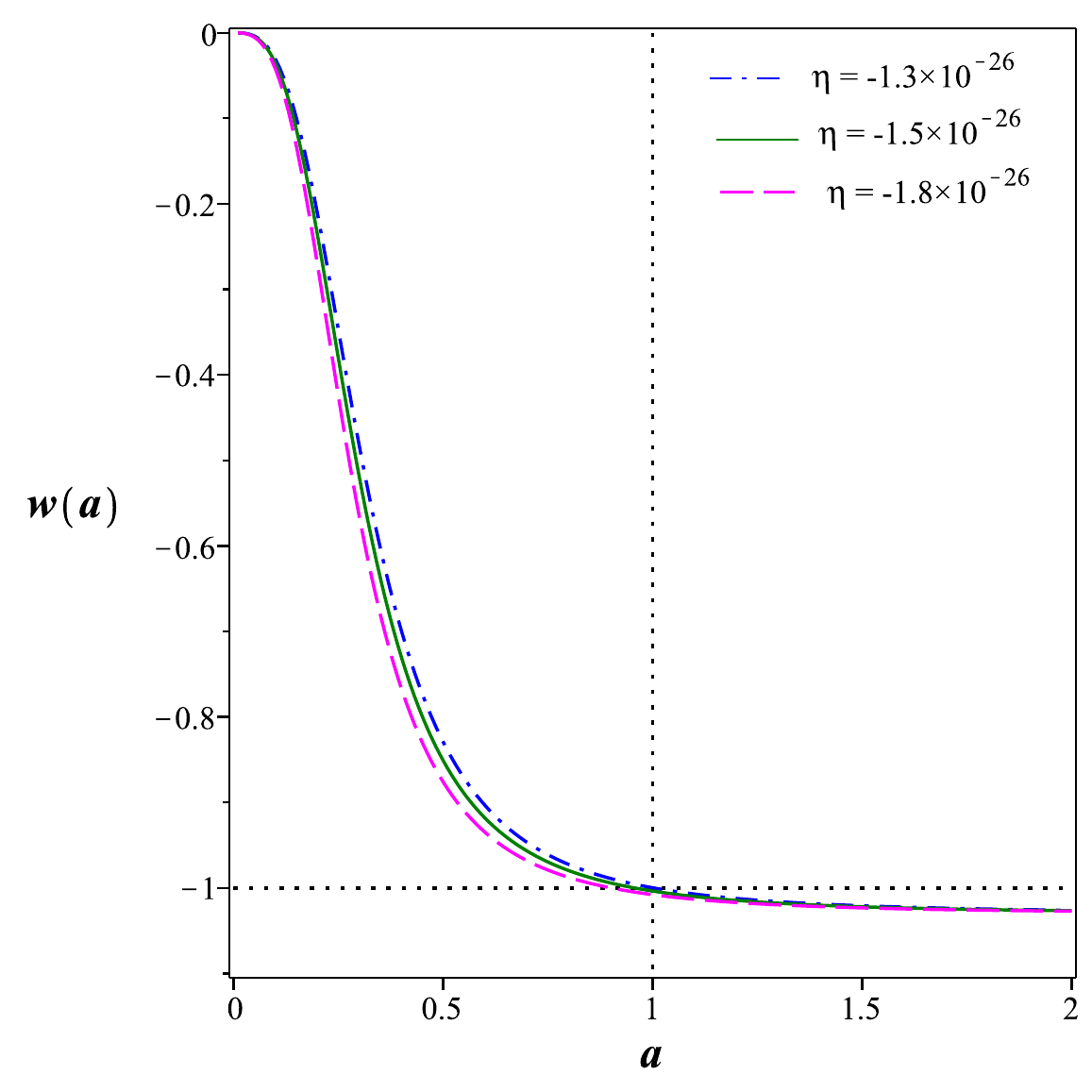}
		\caption{\small Equation of state parameter as a function of scale factor $a$.}
		\label{fig3}
	\end{figure}

	\begin{figure}
		\centering
		\includegraphics[width=0.38\linewidth]{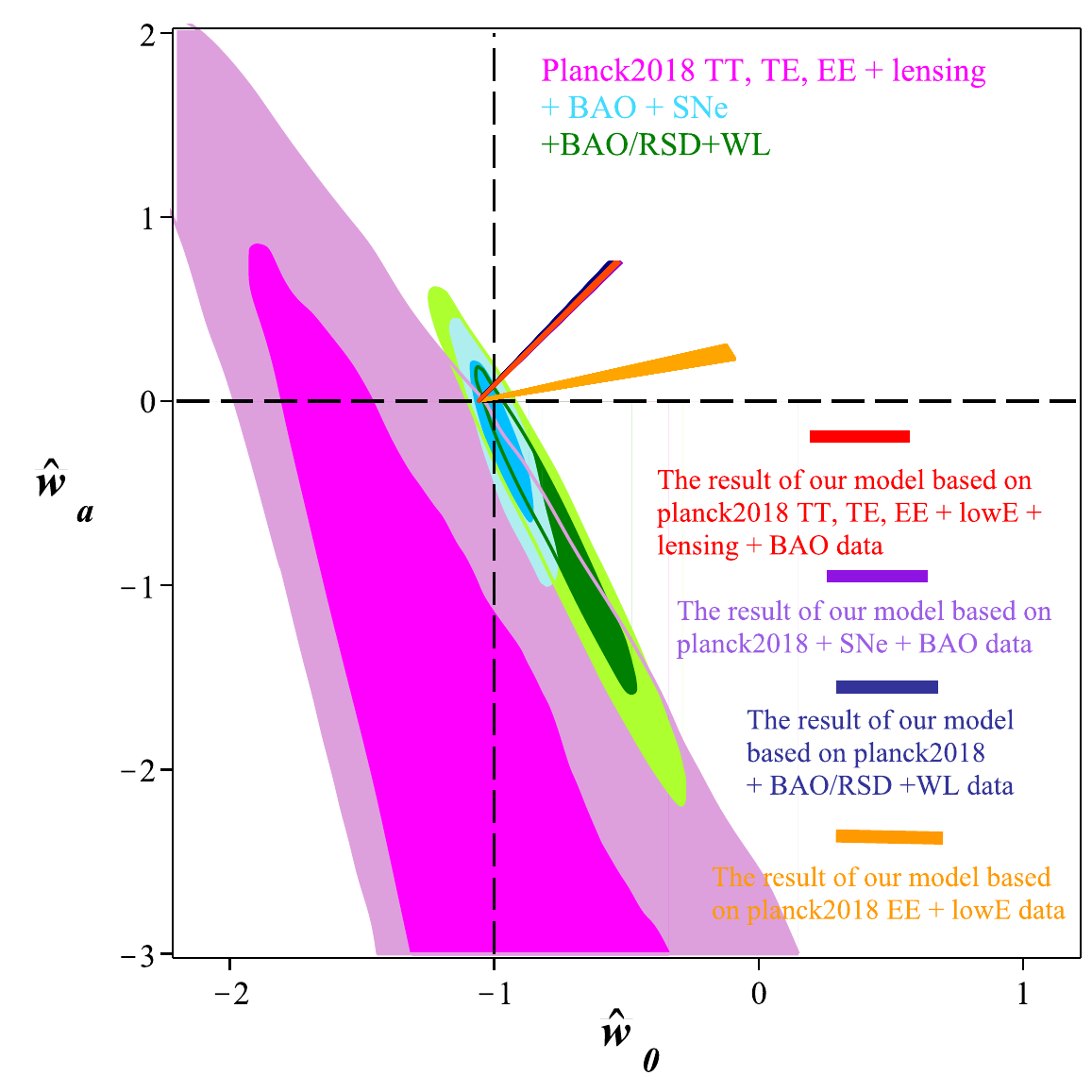}
		\includegraphics[width=0.38\linewidth]{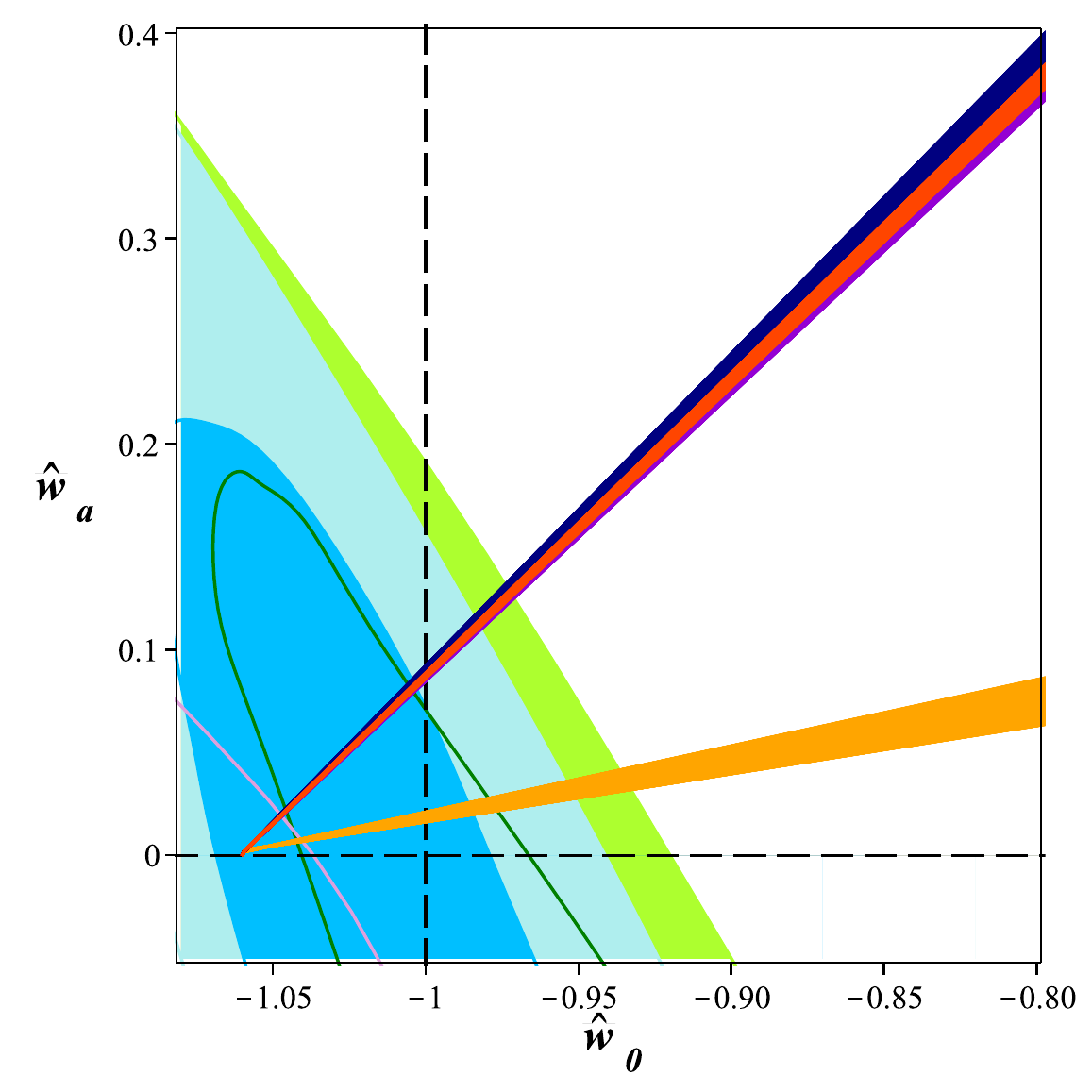}
		\caption{\small Distributions of the $\Big(\hat{w}_0\equiv w(a=1)$,$\hat{w}_a \equiv -\frac{dw}{da}(a=1)\Big)$ parameters for different data combinations contains planck in the EUP background.  Colored markers indicate theoretical values corresponding to different choices of the deformation parameter $\eta$, ranging of the order of $10^{-27}$. The right panel provides a zoomed-in view of the diagram to show the distribution in greater detail.}
		\label{fig4}
	\end{figure}

	\begin{figure}
		\centering
		\includegraphics[width=0.38\linewidth]{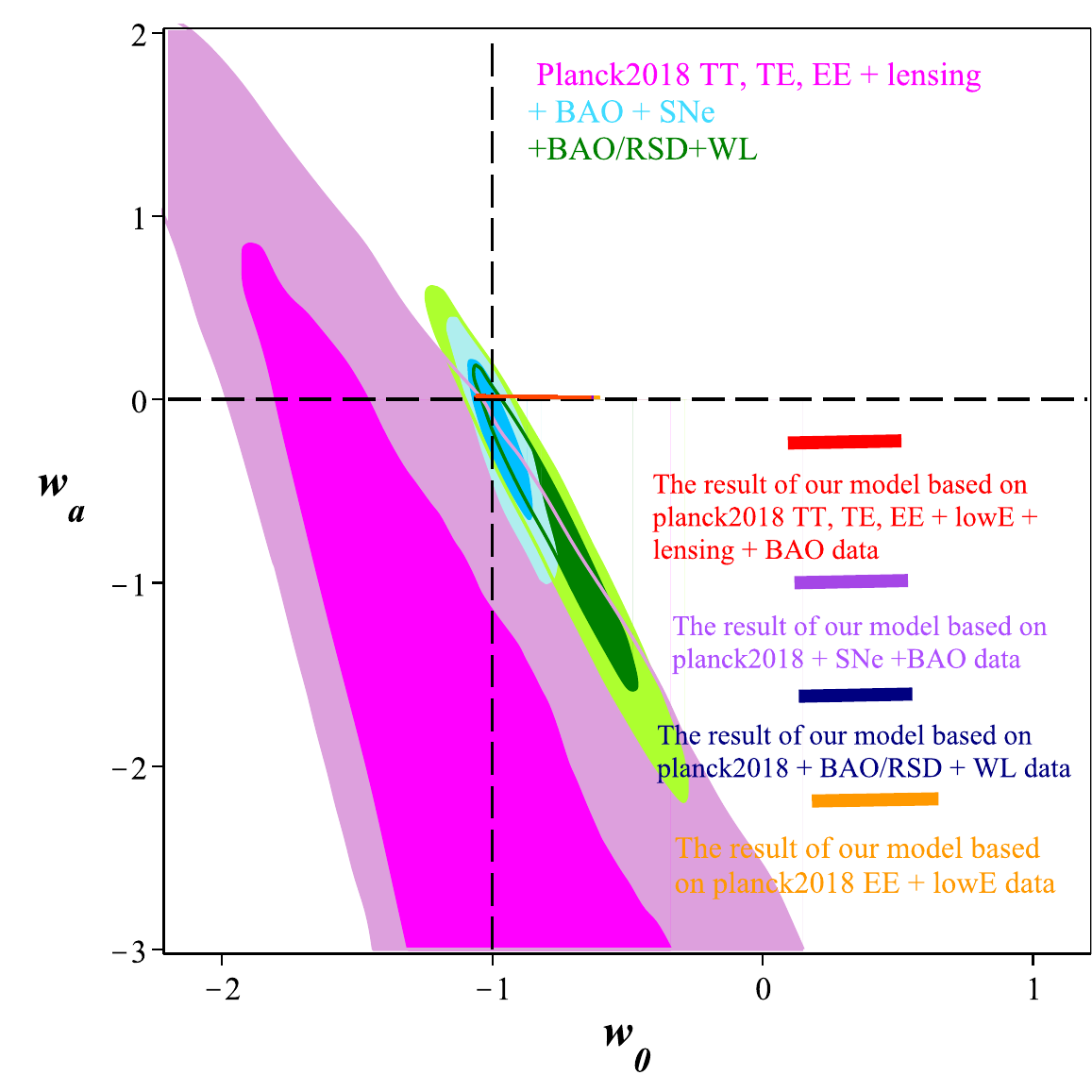}
		\includegraphics[width=0.38\linewidth]{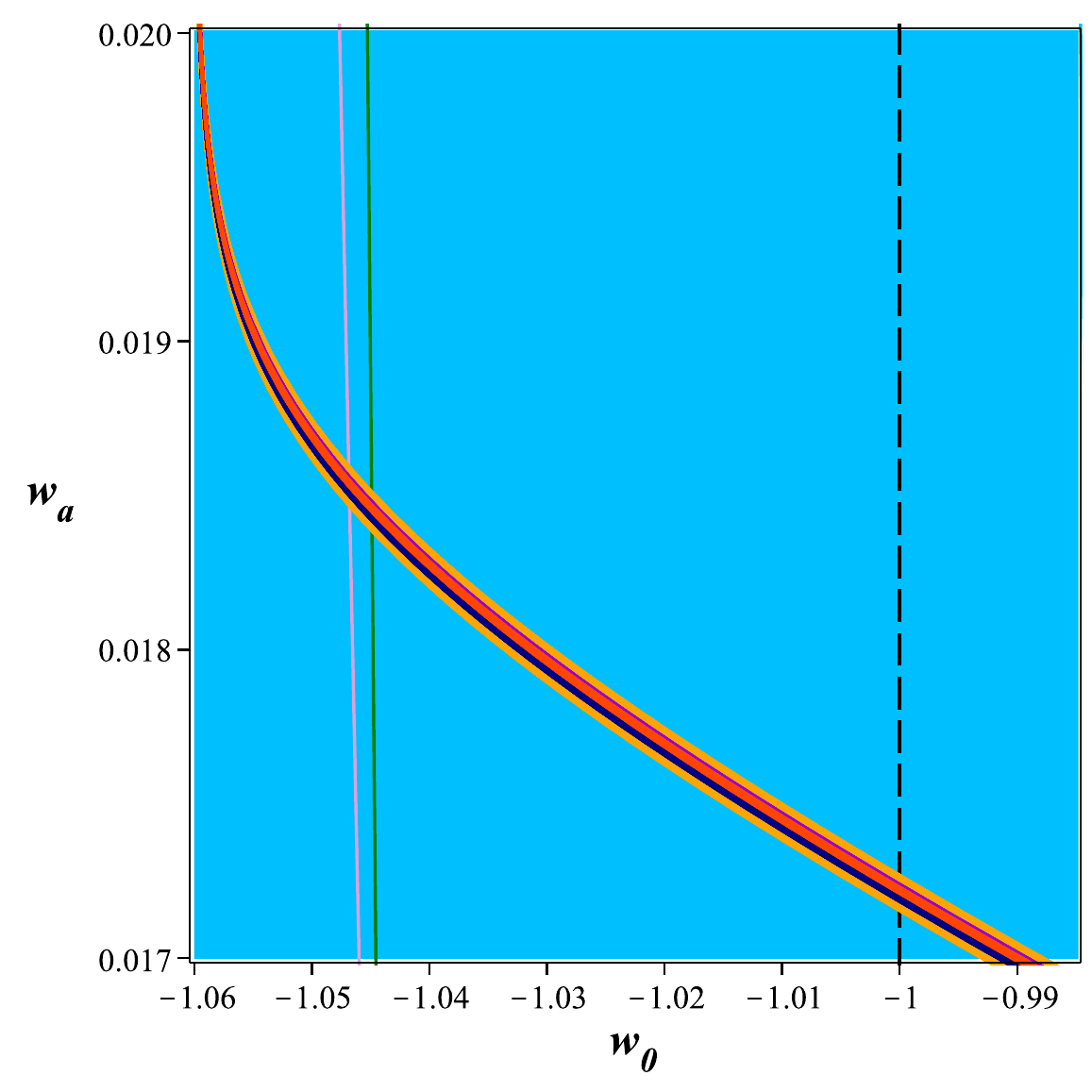}
		\caption{\small Distributions of the $\Big(w_0$, $w_a\Big)$ parameters for different data combinations containing Planck, in the EUP background. Colored markers represent the CPL-equivalent values fitted from the EUP model for different choices of the deformation parameter $\eta$, ranging of the order of $10^{-27}$. The contours correspond to the observational 68\% and 95\% confidence levels from Planck 2018, BAO, and SN data. The right panel provides a zoomed-in view of the diagram to show the parameter space in greater detail.}
		\label{fig5}
	\end{figure}
	
	\begin{figure}
		\centering
		\includegraphics[width=0.35\linewidth]{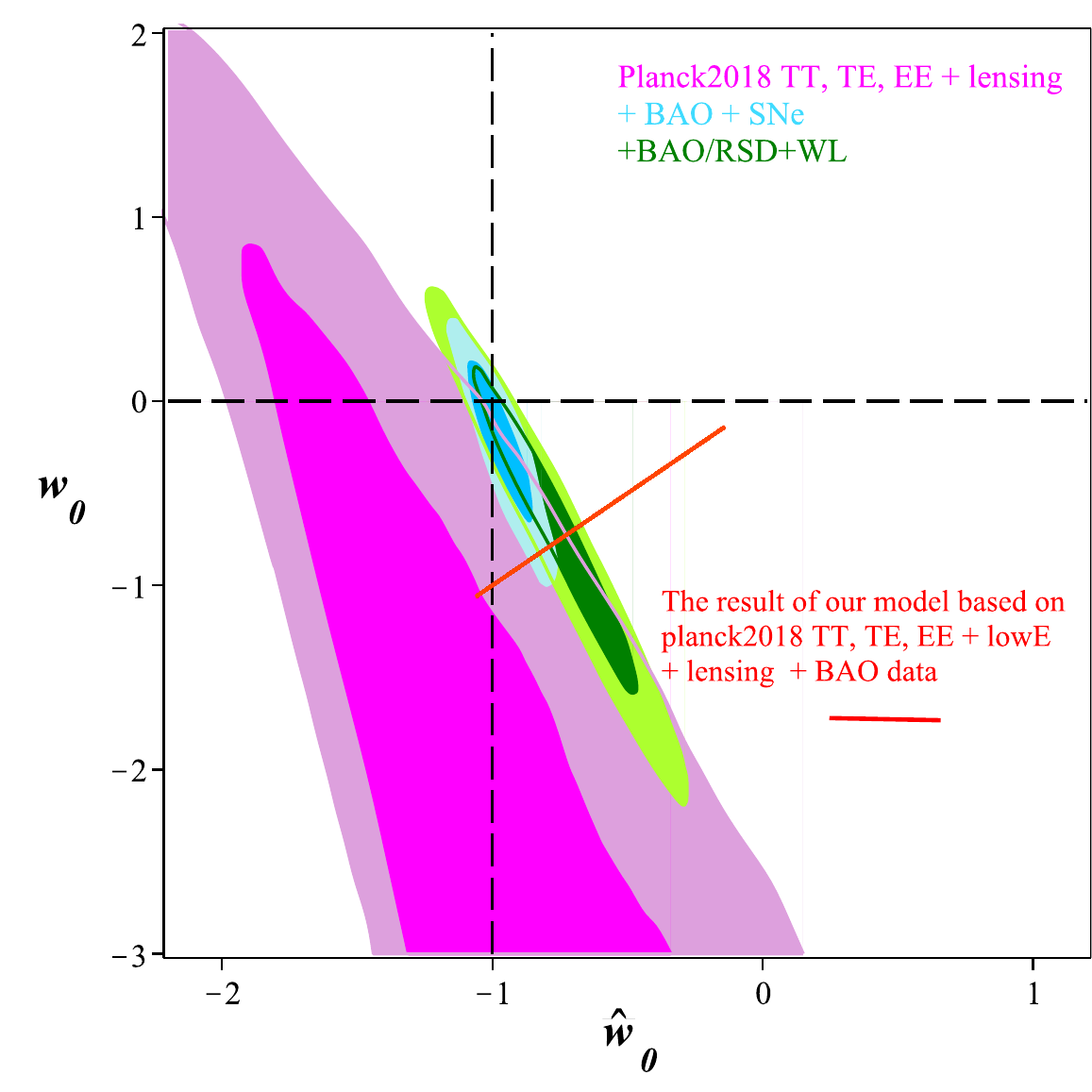}
		\includegraphics[width=0.35\linewidth]{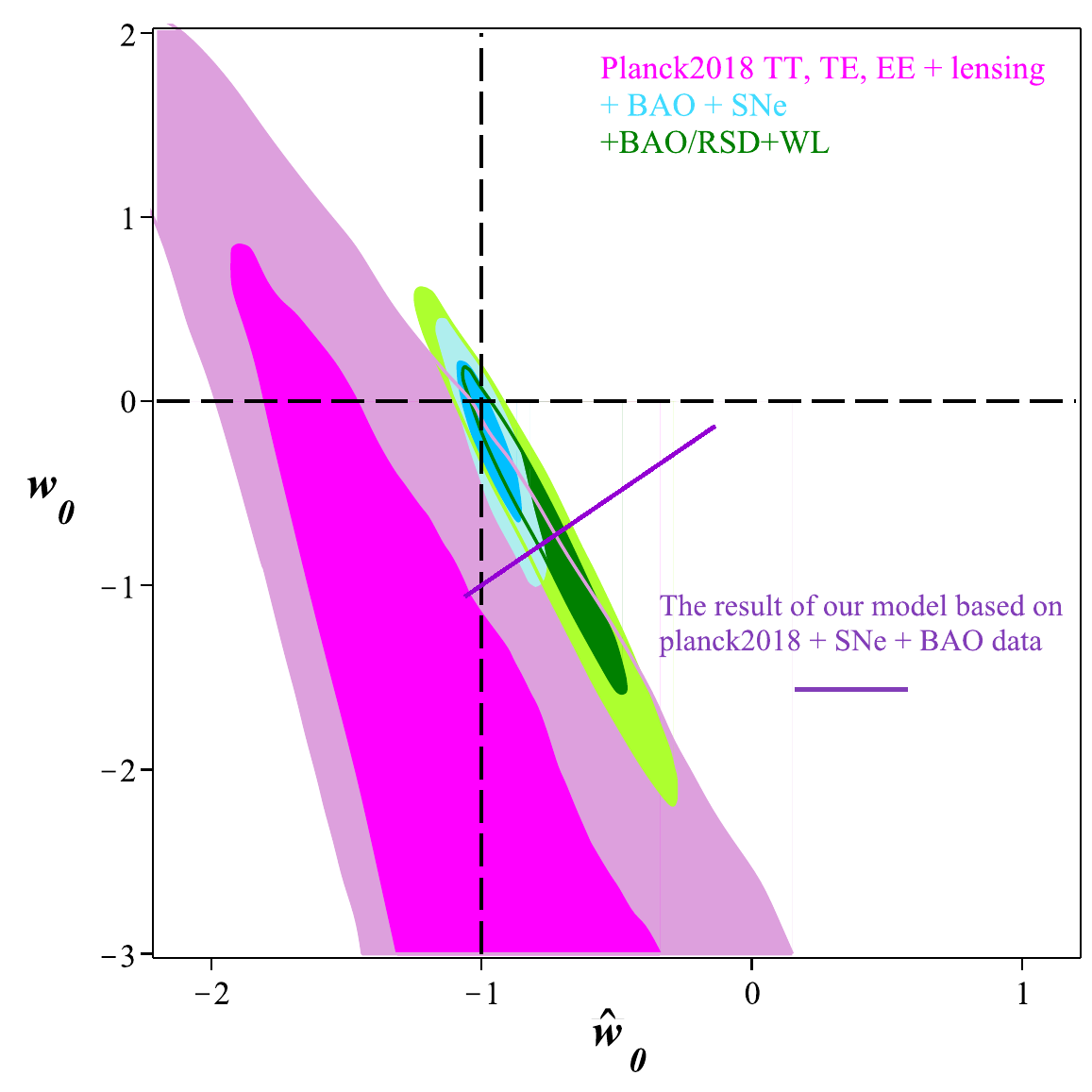}
		\includegraphics[width=0.35\linewidth]{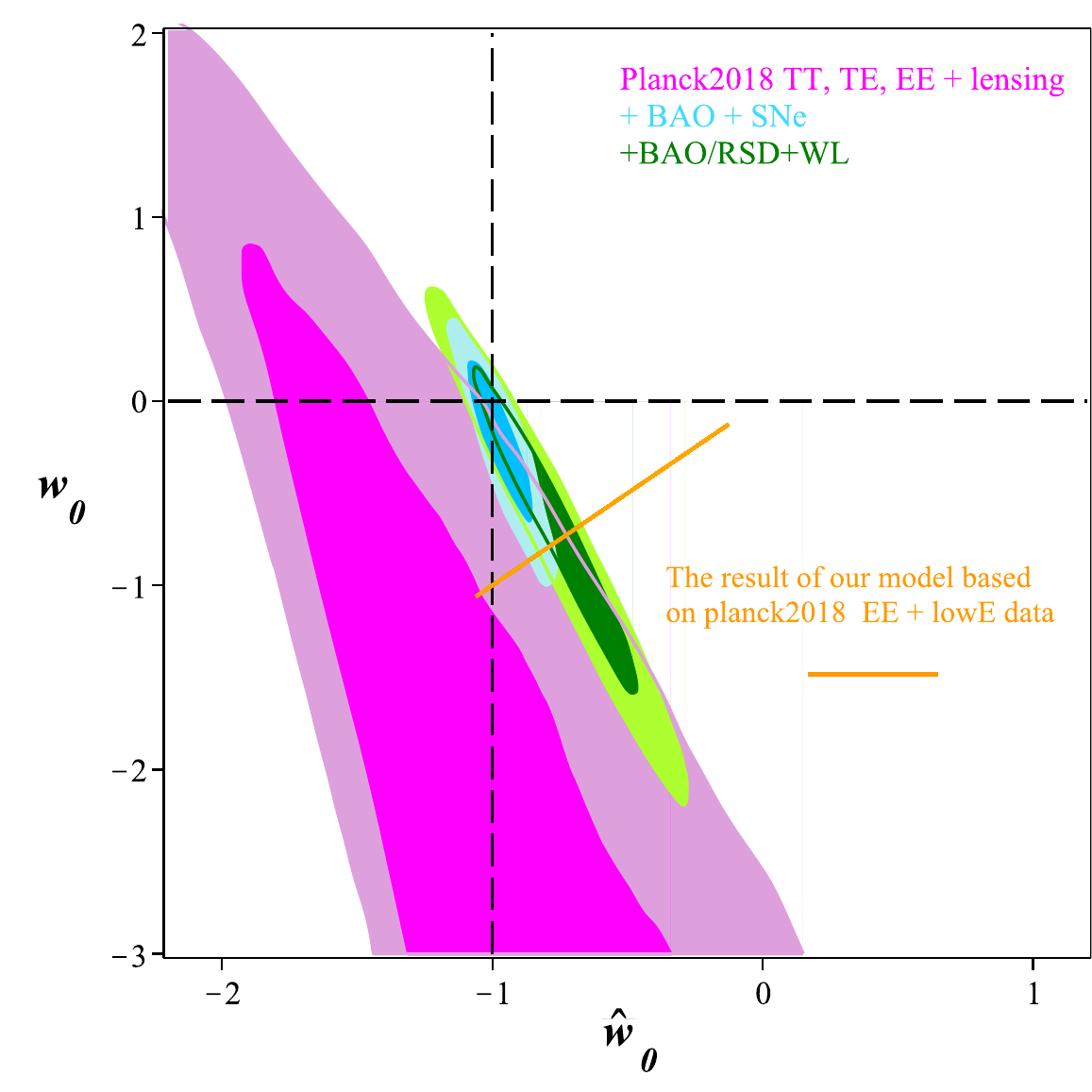}
		\includegraphics[width=0.35\linewidth]{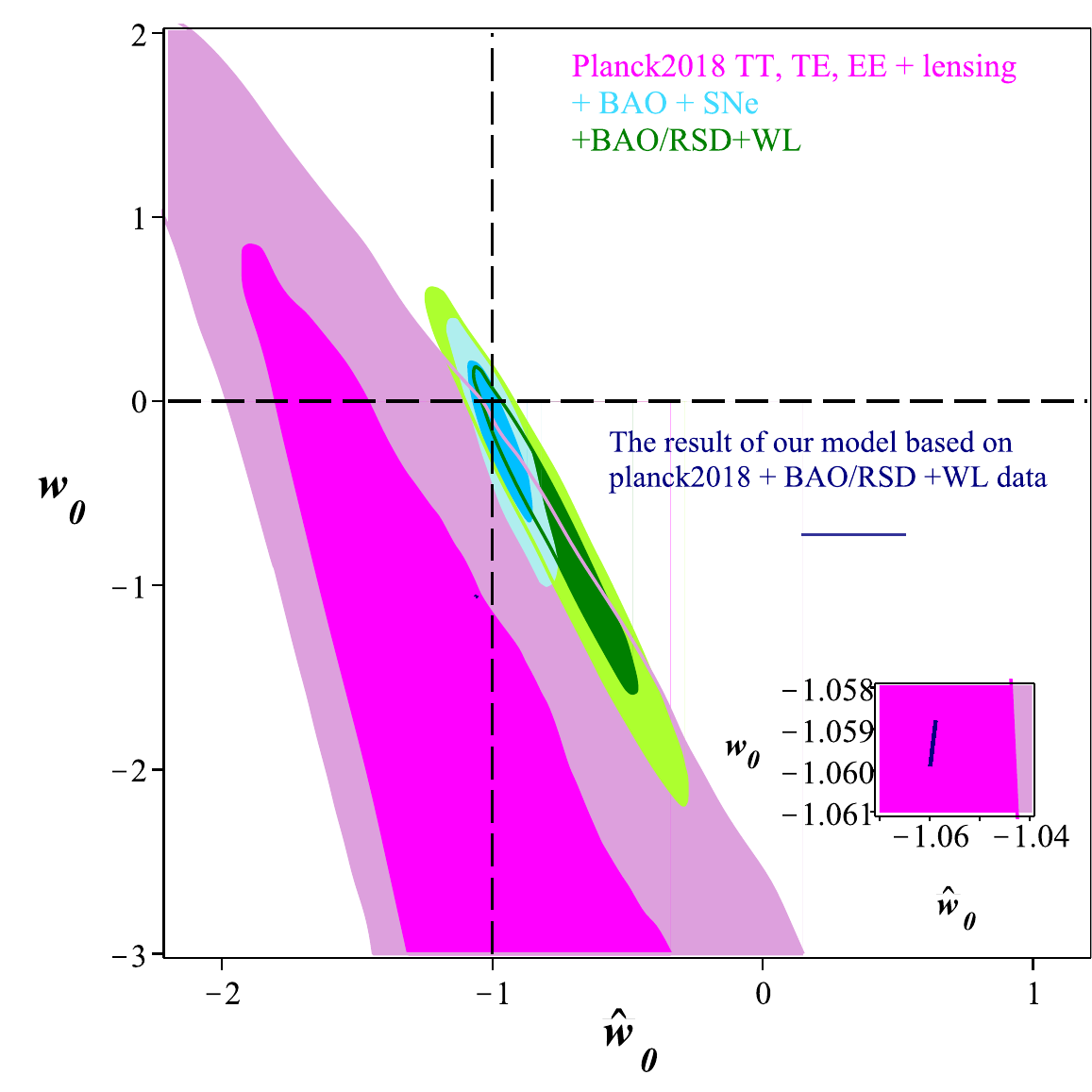}
		\caption{\small Distributions of the $(\hat{w}_0$,${w}_a)$ parameters for different data combinations contains planck in the EUP background.}
		\label{fig6}
	\end{figure}

	\begin{figure}
		\centering
		\includegraphics[width=0.38\linewidth]{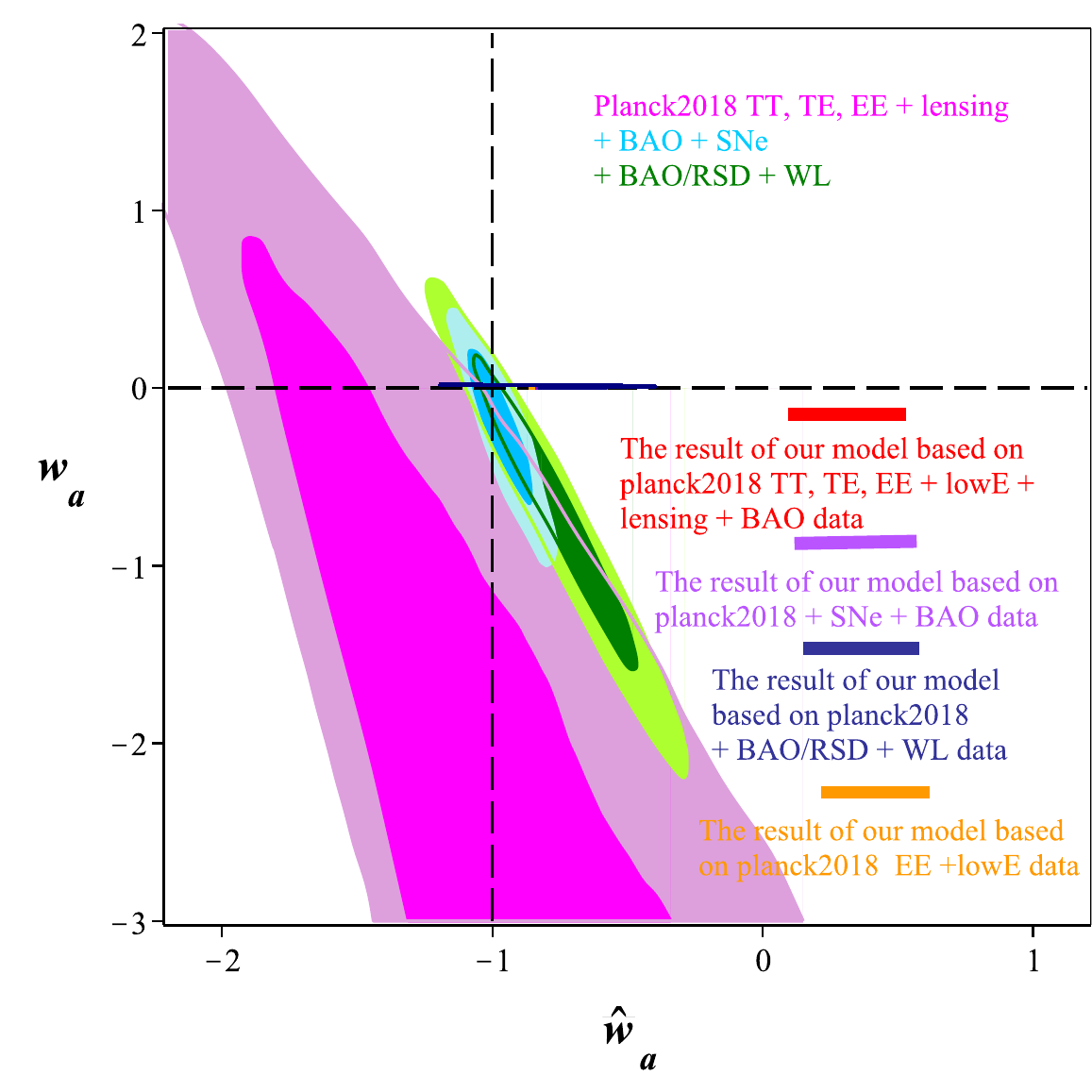}
		\includegraphics[width=0.38\linewidth]{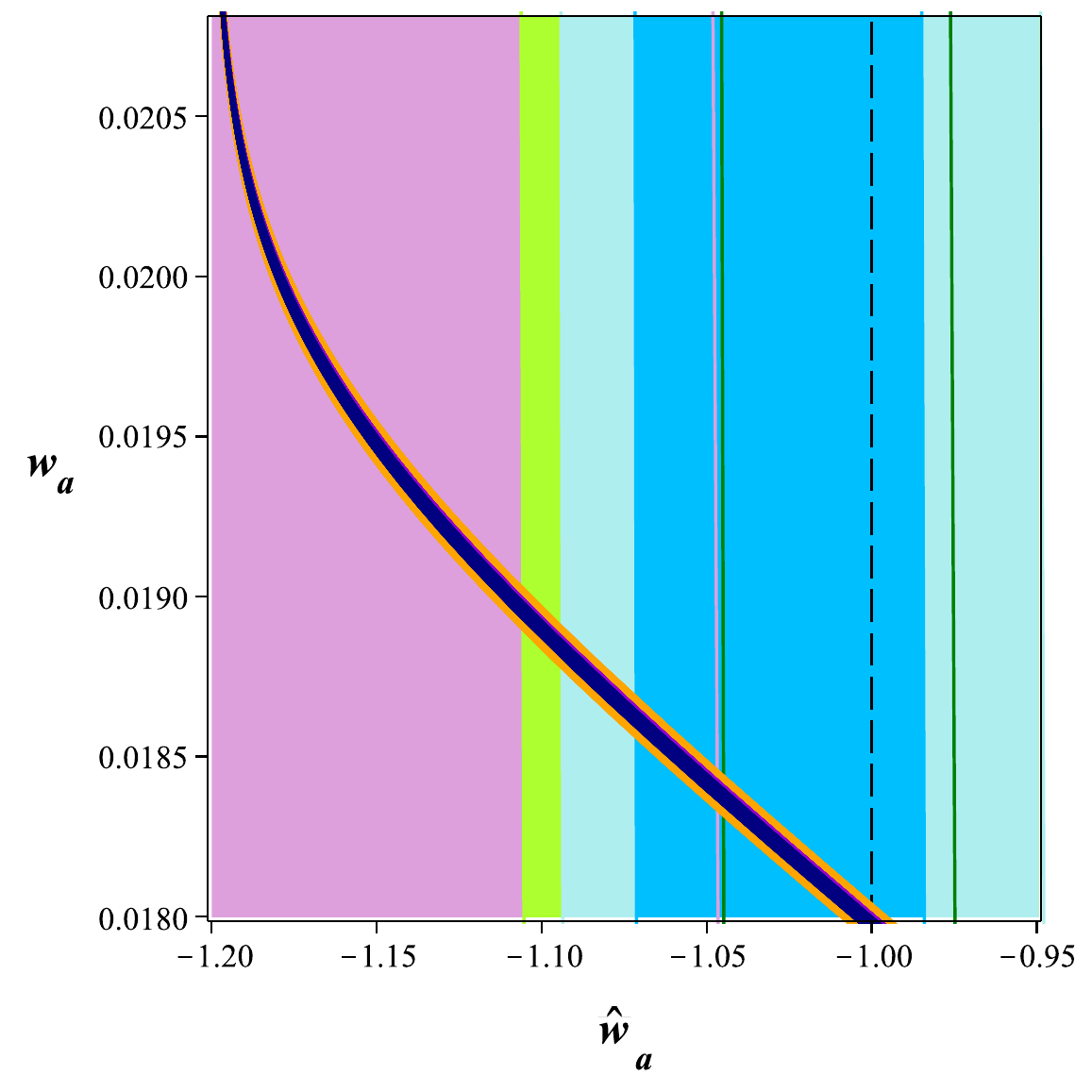}
		\caption{\small Distributions of the $(\hat{w}_a$,${w}_a)$ parameters for different data combinations contains planck in the EUP background. The right panel provides a zoomed-in view for greater clarity near the observationally favored region.}
		\label{fig7}
	\end{figure}

	Figure 3 shows the evolution of the effective equation of state parameter $w(a)$, as predicted by the EUP framework for several values of the negative deformation parameter $\eta<0$. For small values of $|\eta|$, the behavior of $w(a)$ remains close to $-1$, mimicking a cosmological constant. As $|\eta|$
	increases, the model transitions from a quintessence-like regime ($w>-1$) at earlier times to a phantom regime ($w<-1$) at late times. This smooth, monotonic transition arises purely from the infrared structure encoded in the EUP and occurs without introducing scalar fields or modifying general relativity, highlighting the model's natural ability to interpolate between standard and dynamical dark energy scenarios.
	
	Figure 4 presents the theoretical predictions of the EUP model in the $(\hat{w}_0,\hat{w}_a)$ plane, where, $\hat{w}_0=w(a=1)$ is the present-day value and $\hat{w}_a=-\frac{dw}{da}|_{a=1}$ characterizes the evolution of the effective equation of state. These values are derived analytically from the EUP-based expression for $w(a)$ and are overlaid on the marginalized $68\%$ and $95\%$ confidence contours obtained from Planck 2018, BAO, and SN data. The EUP predictions, corresponding to various values of $\eta$ of the order of $10^{-27}$, fall well within the observational bounds, exhibiting a smooth progression from quintessence-like to phantom-like behavior. This alignment supports the consistency of EUP-induced infrared corrections with current cosmological data and suggests a quantum gravitational origin for the observed acceleration.
	
	Figure 5 illustrates the redshift evolution of the effective equation of state $w(a)$ for various values of $\eta$ within the EUP framework. These theoretical predictions are overlaid on the $68\%$ and $95\%$ confidence regions reconstructed from Planck 2018, BAO, and SN data using the CPL parameterization. Across all redshifts, the model curves remain entirely within the observational bounds, exhibiting a transition from $w>-1$ at early times to $w<-1$ near the present epoch. This behavior demonstrates that the EUP model successfully captures key features of dark energy dynamics, without requiring additional fields or free parameters beyond a single, physically motivated deformation scale.
	
	Figure 6 shows the likelihood distribution of the deformation parameter $\eta$, derived from a combined analysis of the Planck 2018, BAO, and SN datasets. The distribution peaks in the negative range, around $\eta \sim \mathcal{O}(10^{-27})$, suggesting a statistical preference for nonzero EUP-induced corrections over the standard case of $\eta = 0$. This best fit range produces effective equation of state trajectories that closely match observational data. The results provide quantitative support for EUP-induced infrared effects as a plausible explanation for late time cosmic acceleration.
	
	Figure 7 presents the posterior distribution of the deformation parameter $\eta$, obtained from the same combination of observational datasets. The shaded regions indicate the $68\%$ and $95\%$ confidence intervals, illustrating the range of $\eta$ values consistent with the data and clearly disfavoring the $\eta = 0$ limit associated with standard cosmology. The posterior peaks around $\eta \sim 10^{-27}$, suggesting that EUP corrections not only align with current observational constraints but may in fact be statistically favored. This finding provides direct bounds on the strength of infrared quantum gravitational corrections and highlights a rare opportunity to probe fundamental physics through cosmological observations.\\
	
	Having established the consistency of the EUP-modified cosmological model with current observational constraints in the $(w_0, w_a)$ parameter space, it is now essential to examine the physical viability of the resulting cosmic fluid beyond its kinematic behavior. In particular, the compatibility of the model with classical energy conditions (such as the null, weak, strong, and dominant energy conditions) provides critical insight into the underlying causal structure, geodesic focusing, and energy flow properties of the effective energy-momentum tensor induced by the EUP corrections. These conditions serve as important diagnostic tools in general relativity and cosmology, offering theoretical constraints that any viable modification of gravity or exotic matter scenario must satisfy. In the following section, we analyze the extent to which the EUP-based model adheres to these energy conditions, shedding further light on its physical robustness and gravitational consistency.
	
	\section{Energy Conditions}
	
	To assess the physical admissibility of the cosmic fluid in the EUP-modified cosmology, we evaluate its compliance with the standard energy conditions of general relativity. These include the null energy condition (NEC), weak energy condition (WEC), strong energy condition (SEC), and dominant energy condition (DEC). Each imposes specific constraints on the effective energy density $\rho(z)$ and pressure $p(z)$, as defined by the EUP-modified Friedmann dynamics (see Equations~(14) and (15)).
	
	We compute the relevant combinations over redshift in the range $-1 \leq z \leq 3$, assuming a representative value of the deformation parameter $\eta =-3\times 10^{-27}$. The results, shown in Figure 8, demonstrate the model’s consistency with physically reasonable energy propagation and provide insight into its late-time acceleration behavior.
	
	\subsection{Null Energy Condition (NEC)}
	
	The NEC requires $\rho + p \geq 0$. As illustrated in Figure 8, this condition is satisfied for all redshifts considered, with $\rho + p$ approaching zero at low redshift. This ensures that the model does not violate causal propagation of energy or gravitational focusing constraints.
	
	\subsection{Weak Energy Condition (WEC)}
	
	The WEC requires that the energy density is non-negative and that $\rho \geq 0$ and $\rho + p \geq 0$. Our numerical results confirm that $\rho(z)$ remains strictly positive across all redshifts, and, consistent with the NEC, that $\rho + p \geq 0$. The WEC is therefore satisfied throughout.
	
	\subsection{Strong Energy Condition (SEC)}
	
	The SEC requires both $\rho + p \geq 0$ and $\rho + 3p \geq 0$.While the first condition is satisfied as discussed above, we find that $\rho + 3p$ becomes negative at low redshifts (see Figure 8). This violation of the SEC is a necessary feature for the onset of cosmic acceleration and is a common trait of accelerating cosmologies.
	
	\subsection{Dominant Energy Condition (DEC)}
	
	The DEC imposes $\rho \geq 0$ and $\rho \geq |p|$, ensuring causal energy flow and the non-exotic character of the effective fluid. Our analysis confirms that the inequality \(\rho \geq |p|\) holds for all redshifts in the relevant range. Thus, the DEC is also satisfied.

	\begin{figure}
		\centering
		\includegraphics[width=0.38\linewidth]{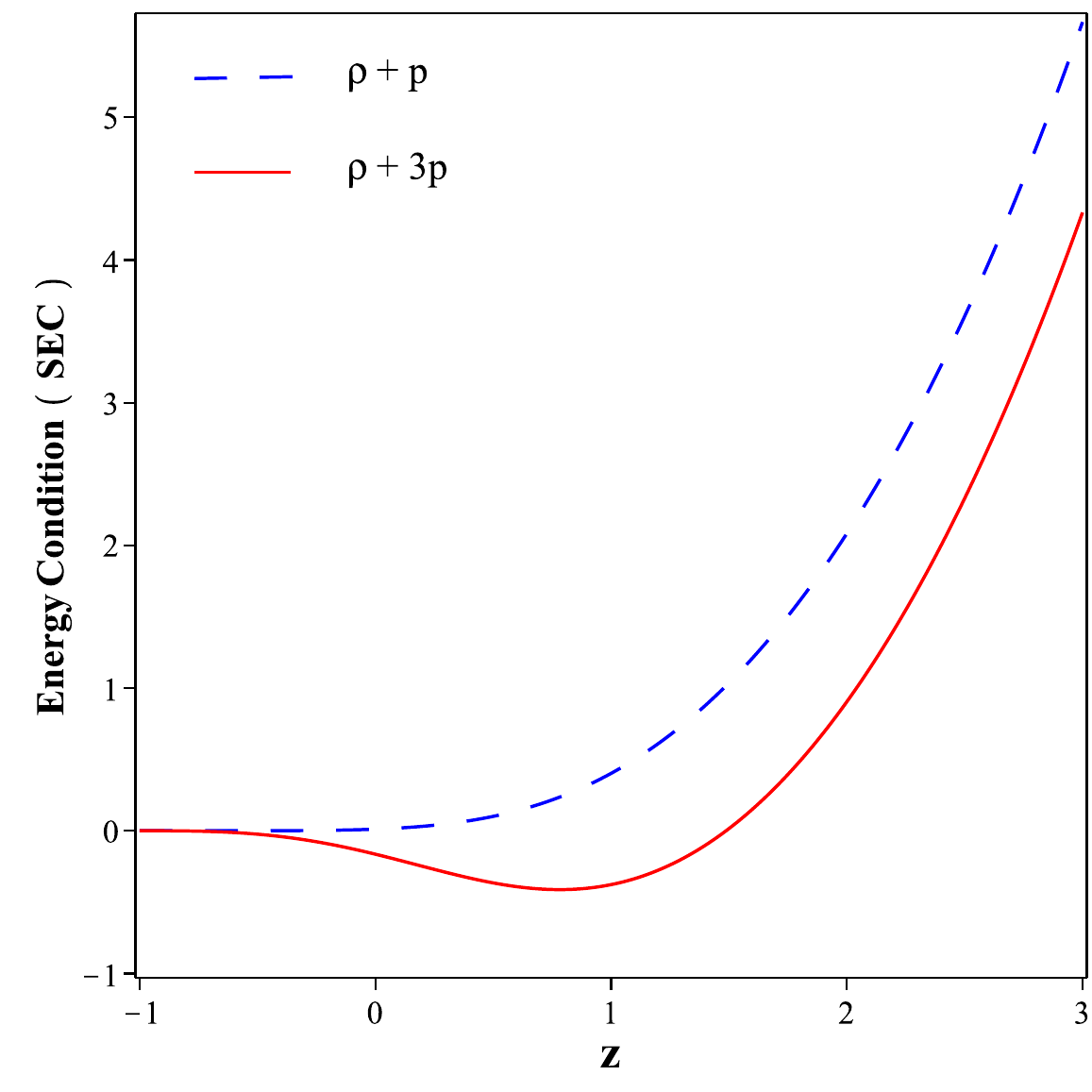}
		\includegraphics[width=0.38\linewidth]{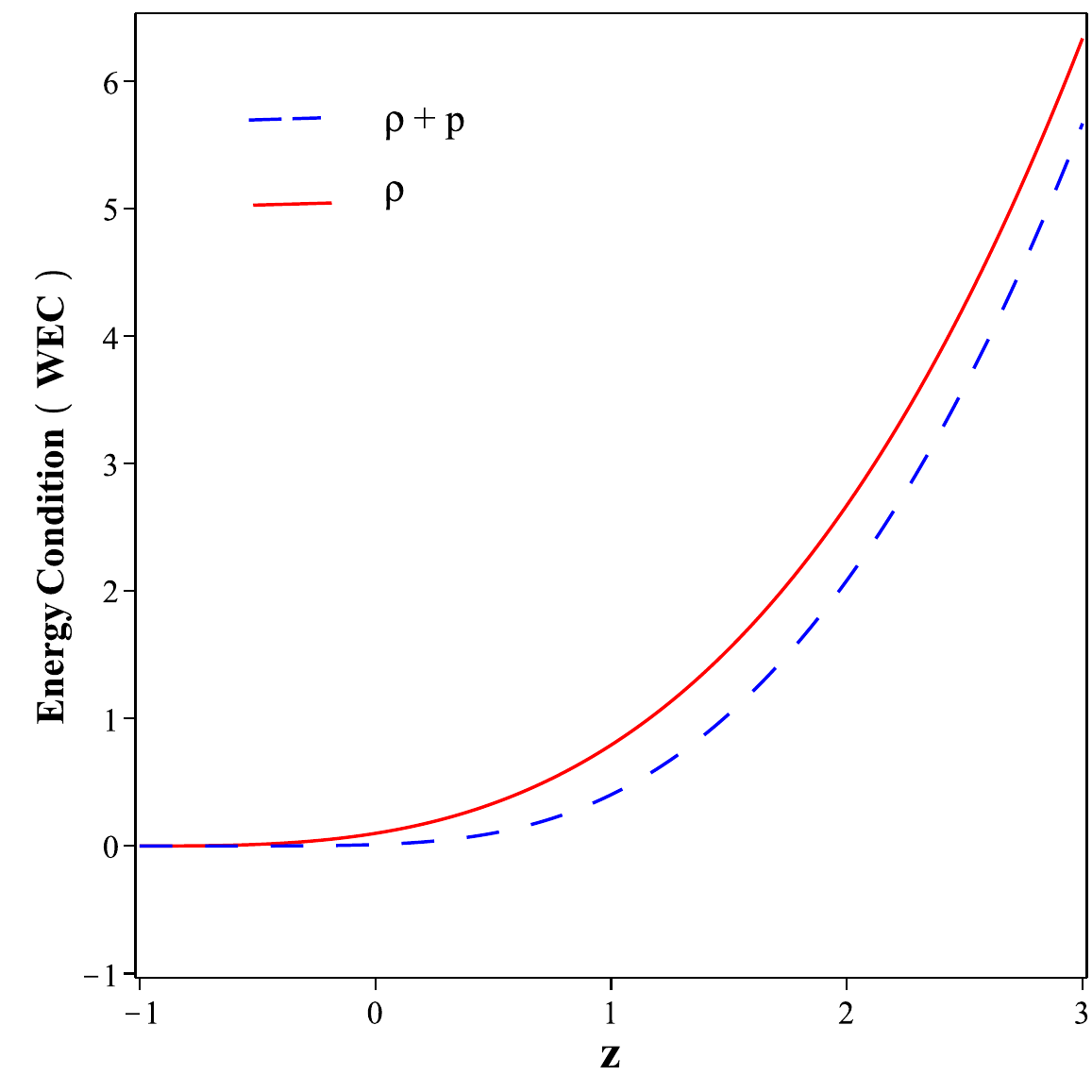}
		\includegraphics[width=0.38\linewidth]{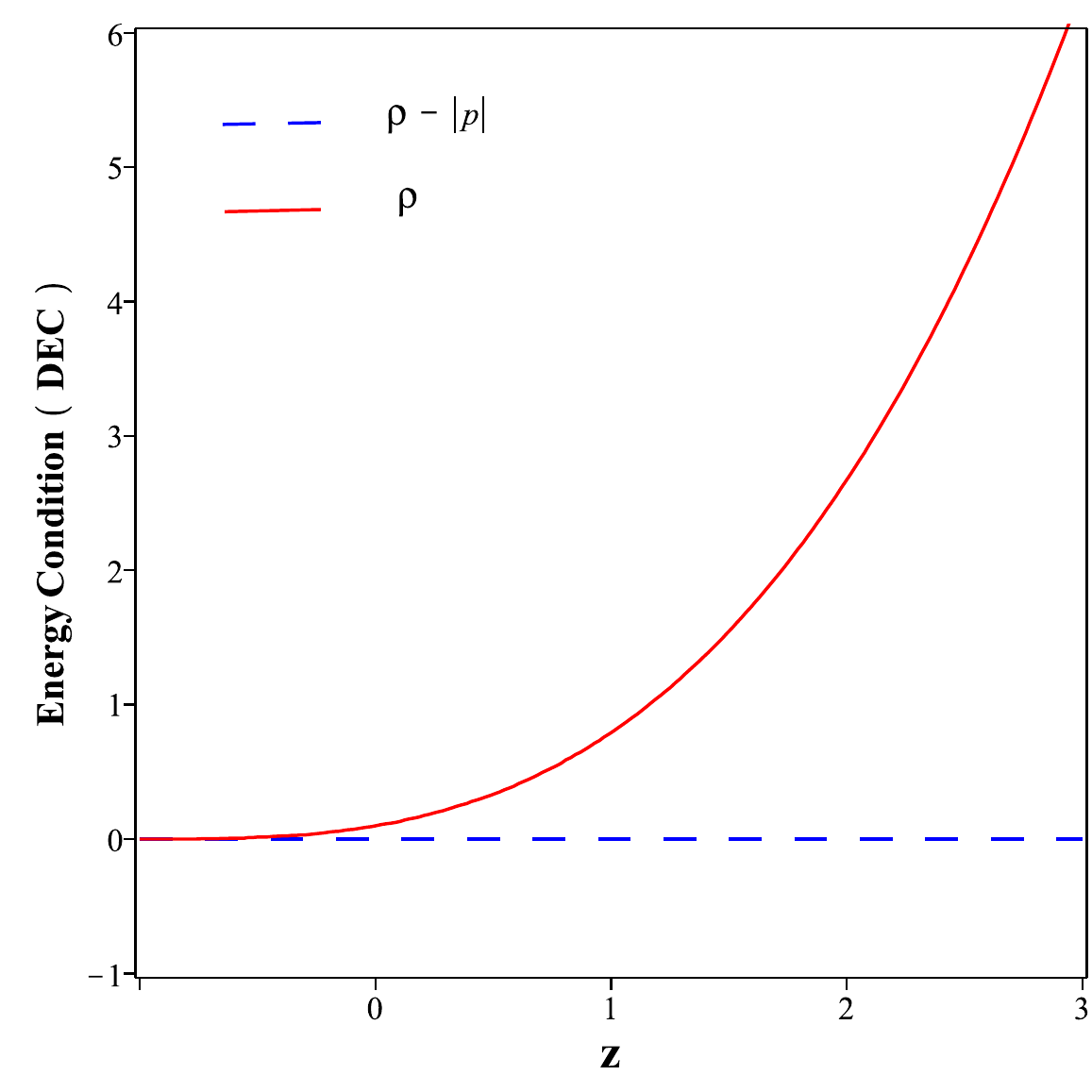}
		\includegraphics[width=0.38\linewidth]{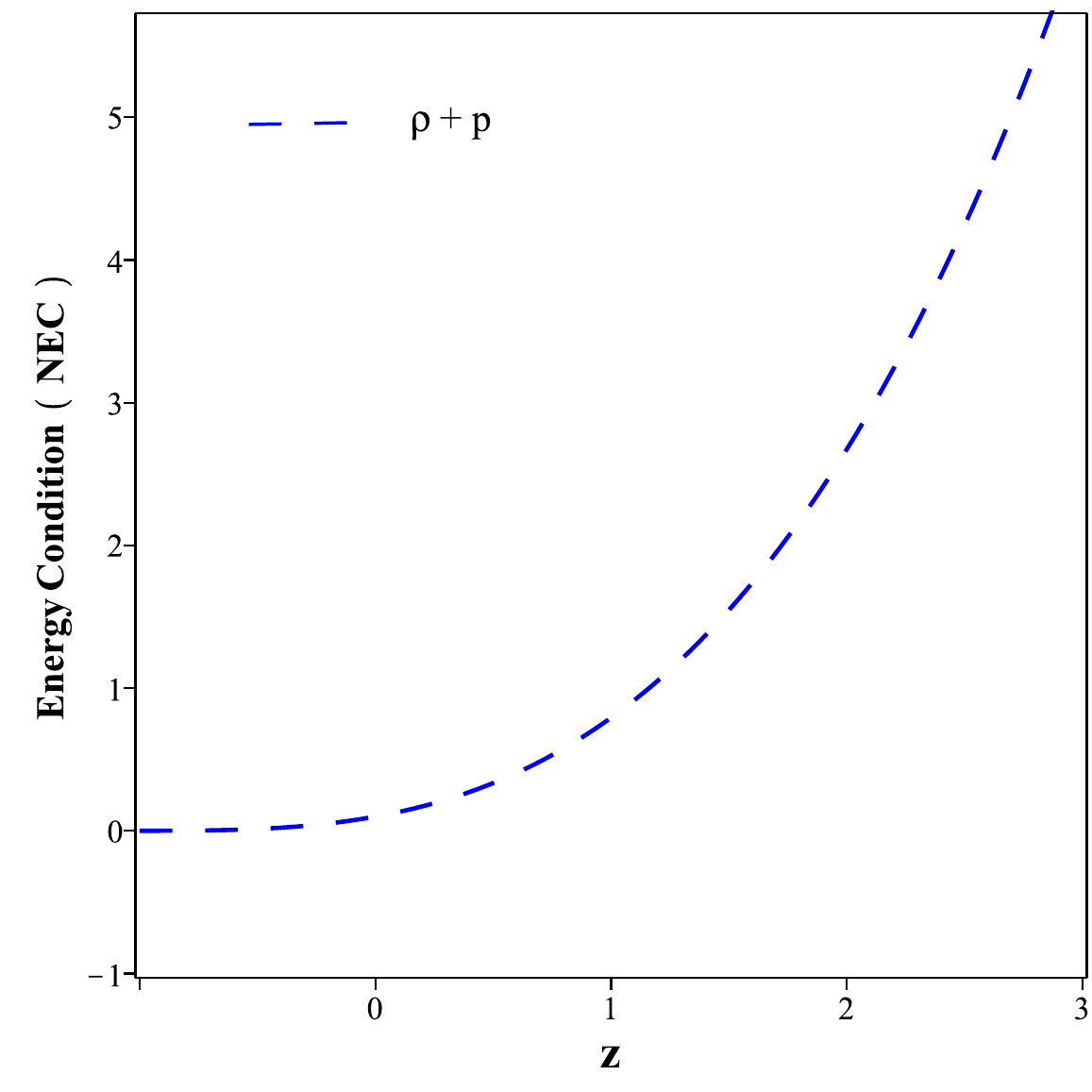}
		\caption{\small Evolution of energy condition quantities in the EUP-corrected cosmology, for \(\eta = -3\times10^{-27}\). Shown are the energy density \(\rho\), and the combinations \(\rho + p\), \(\rho + 3p\), and \(\rho - |p|\). NEC, WEC, and DEC are satisfied across the redshift range, while SEC is violated at late times, consistent with accelerated expansion.}
		\label{fig8}
	\end{figure}

	This pattern of energy condition behavior (in which NEC, WEC, and DEC are satisfied and SEC is violated) is characteristic of viable dark energy models. It confirms that the infrared deformation introduced by the EUP acts as a physically plausible source of late-time cosmic acceleration without requiring exotic matter content.

	\section{Dynamical and Thermodynamic Stability}
	
	In addition to consistency with observational data and classical energy conditions, a viable cosmological model must exhibit both dynamical and thermodynamic stability. In this section, we analyze the EUP-corrected cosmology from this perspective. First, we examine the phase space behavior using a dynamical systems approach. Then, we evaluate the effective sound speed to assess causal propagation and potential instabilities in perturbations.
	
	\subsection{Phase Space and Late Time Behavior}
	
	As discussed in Section 3, the EUP correction leads to an effective equation of state
	$w(a) \to -1$ at late times, indicating that the EUP-induced fluid asymptotically behaves
	like a cosmological constant. This corresponds to the de Sitter limit with $p = -\rho$
	and a constant Hubble parameter $H \to H_0$. Such asymptotic behavior suggests the
	presence of a stable fixed point in the dynamical system, toward which the cosmological
	evolution naturally flows.
	
	To characterize this evolution more precisely, we introduce the dimensionless variables
	
	\begin{eqnarray}
		\label{eq22}x \equiv \frac{\rho_m}{3H^2}, \qquad y \equiv \frac{\rho_{\text{EUP}}}{3H^2},
	\end{eqnarray}
	which represent the normalized energy densities of matter and the EUP-induced component, respectively. The total Friedmann constraint implies $ x + y = 1 $ in the absence of additional species.
	
	The autonomous system governing their evolution with respect to the e-folding time $ N = \ln a $ is given by
	\begin{eqnarray}
		\label{eq23}\frac{dx}{dN} &= -3x(1 - x) - 3xy(1 + \omega_{\text{EUP}})\,,
	\end{eqnarray}
	
	\begin{eqnarray}
		\label{eq24}\frac{dy}{dN} &= 3xy - 3y(1 + \omega_{\text{EUP}})(1 - y)\,,
	\end{eqnarray}
	where \( \omega_{_\text{EUP}} \equiv p_{_\text{EUP}}/\rho_{_\text{EUP}} \) is the EUP-induced equation of state parameter.
	
	The system admits critical points at $ (x, y) = (1, 0) $, corresponding to a matter-dominated phase, and at $ (x, y) = (0, 1) $, representing a universe dominated by EUP corrections. Linearizing around these points reveals that the matter-dominated phase is a saddle, while the EUP-dominated phase is a stable attractor. This confirms that the cosmic dynamics naturally evolve toward an accelerating regime sourced by the EUP component.
	
	Figure 9 illustrates the trajectories in the $ (x, y) $ plane, demonstrating this dynamical flow from a decelerating to an accelerating epoch.

	\begin{figure}
		\centering
		\includegraphics[width=0.40\linewidth]{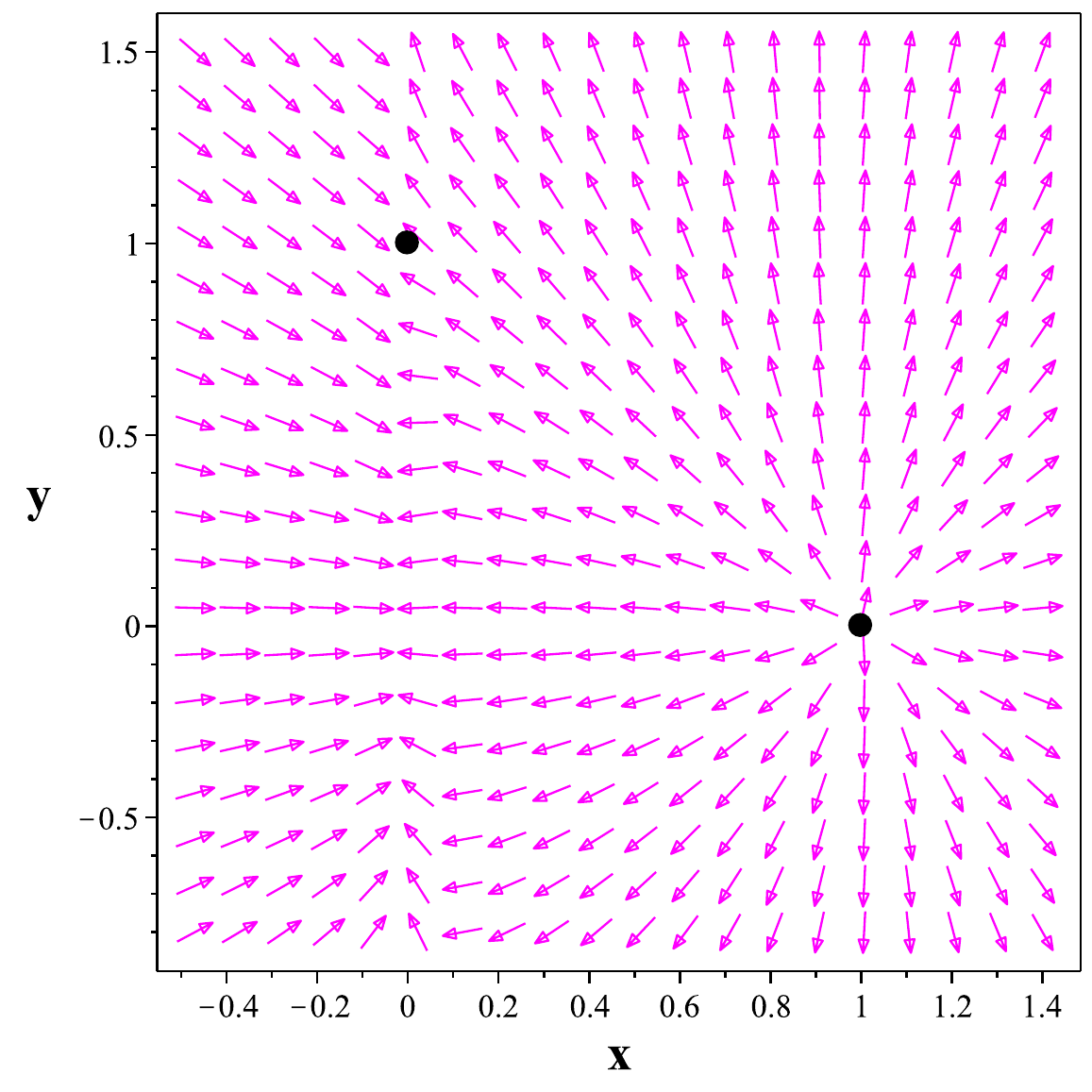}
		\caption{\small Phase space evolution of the EUP-corrected cosmological model in the \( (x, y) \) plane. Trajectories illustrate the transition from a matter-dominated epoch (\( x = 1, y = 0 \)) to a stable EUP-dominated de Sitter phase (\( x = 0, y = 1 \)). The stable attractor behavior confirms the model’s capacity to drive late-time acceleration.}
		\label{fig9}
	\end{figure}
	
	\subsection{Thermodynamic Stability: Sound Speed Analysis}
	
	Thermodynamic stability of a cosmological fluid is commonly assessed through the square of its effective sound speed, defined as
	\begin{eqnarray}
		\label{eq25}c_s^2 \equiv \frac{d p}{d \rho}.
	\end{eqnarray}
	In the context of the EUP-modified Friedmann equations, the expressions for pressure and energy density lead to a nontrivial but analytically tractable form of $c_s^2$. Explicitly, the sound speed squared is given by
	\begin{eqnarray}
		\label{eq26}c_s^2 = -\frac{
			\left( \pi l^2 \rho\, \tilde{\mathcal{W}} + \frac{3}{2} \eta \tilde{\mathcal{W}}^2 - \frac{3}{2} \eta \right)}{
			12 \eta^2 (1 + \tilde{\mathcal{W}})(\tilde{\mathcal{W}} - 1)^2
		}\times\left[ -16 \eta \ln\left( H + \sqrt{H^2 - 16\eta} \right) + H \left( H + \sqrt{H^2 - 16\eta} \right) \right]\,,
	\end{eqnarray}
	where,
	\begin{eqnarray}
		\label{eq27}\tilde{\mathcal{W}} \equiv \mathcal{W}\left(
		-\frac{1}{16\eta} \exp\left( -\frac{2\pi l^2 \rho - 3\eta}{3\eta} \right)
		\right)\,.
	\end{eqnarray}
	
	We numerically evaluate this expression over the redshift interval $ 0 \leq z \leq 3 $ for a representative deformation parameter $ \eta$ of the order of $10^{-27} $. The resulting evolution of $c_s^2(z)$ is shown in Figure~10.

	\begin{figure}
		\centering
		\includegraphics[width=0.44\linewidth]{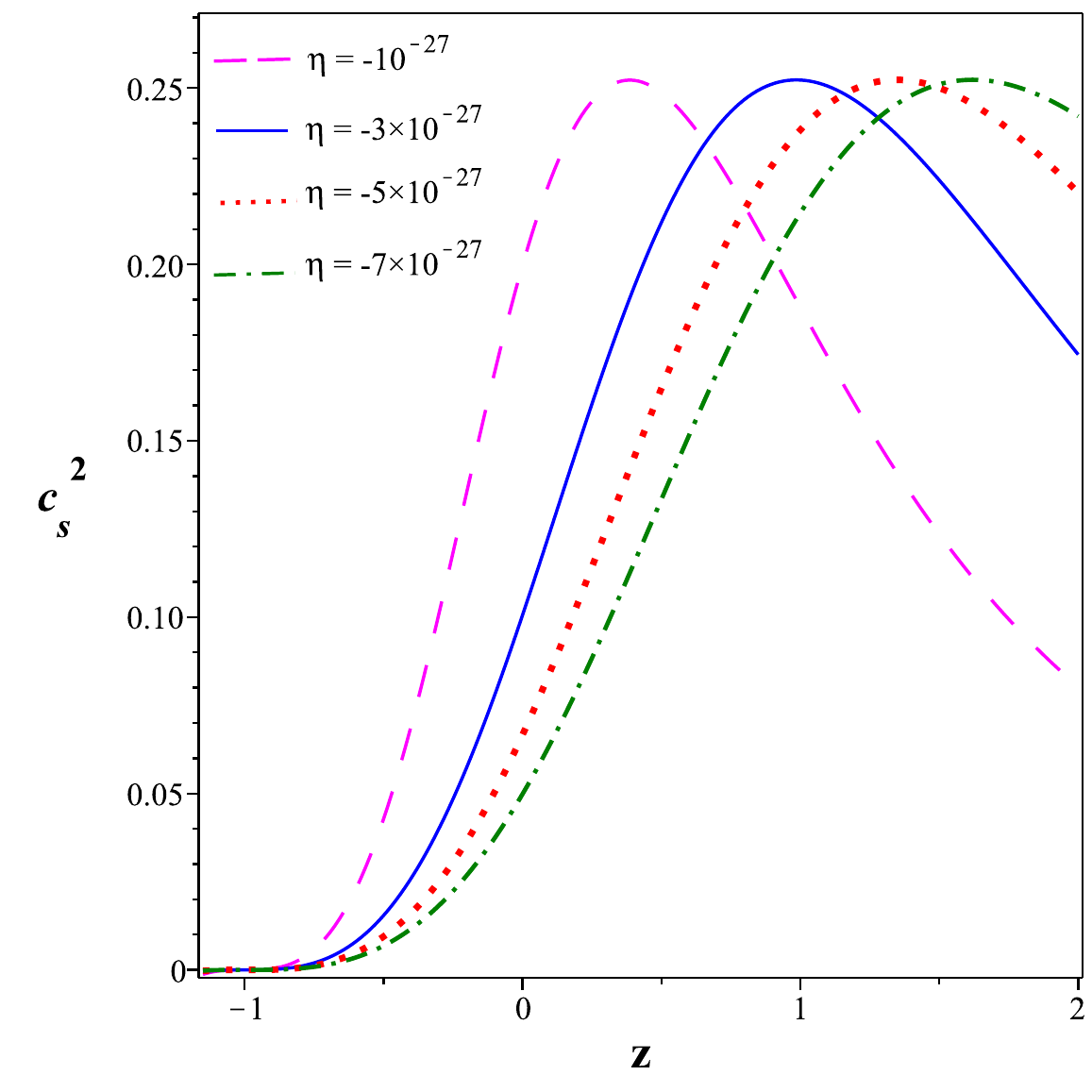}
		\caption{\small Evolution of the effective sound speed squared $c_s^2$ in the EUP-corrected cosmological model for different values of $\eta$. The smooth and finite behavior of the curve reflects the model’s perturbative stability across the redshift range.}
		\label{fig10}
	\end{figure}
	
	As shown in Figure 10, the effective sound speed squared remains positive and evolves smoothly with redshift. This behavior indicates that the EUP-induced effective fluid is free from classical instabilities, such as gradient or ghost modes. The EUP framework thus naturally supports stable perturbation propagation.
	
	This result reinforces the physical viability of the model and supports its interpretation as a stable infrared modification of general relativity that does not require exotic scalar fields or ad hoc matter components.

	\section{Conclusion}
	
	In this work, we explored the cosmological consequences of the Extended Uncertainty Principle (EUP) through its thermodynamic imprint on Friedmann dynamics. By incorporating the EUP into the Clausius relation at the apparent horizon, we derived a modified entropy-area relation and obtained a deformed Friedmann equation with an effective infrared correction term characterized by a negative deformation parameter $\eta$.
	
	The analysis demonstrated that the EUP correction leads to an emergent cosmic fluid with features similar to dark energy. For suitable values of $\eta < 0$, the model exhibits a transition from deceleration to acceleration, consistent with observational evidence. The effective equation of state $w(z)$ crosses the phantom divide line without invoking scalar fields, indicating that the EUP-induced fluid behaves as a unified dark sector component.
	
	Notably, the effective equation of state parameter associated with the EUP fluid asymptotically approaches \(w \approx -1\). This places the model within current observational constraints. In particular, the Planck 2018 data set, combined with other probes, places the dark energy equation of state at $w_0 = -1.03 \pm 0.03$ at $68\%$ confidence level, with our result lying well within the allowed region of the $w_0$-$w_a$ plane.
	
	We tested the energy conditions and found that while the NEC, WEC, and DEC are satisfied across the cosmic timeline, the SEC is violated at late times (a condition required for cosmic acceleration. This pattern is a hallmark of dark energy) driven dynamics and confirms the physical plausibility of the model.
	
	Regarding thermodynamic stability, we evaluated the square of the effective sound speed, $c_s^2 = dp/d\rho$, using the exact expressions derived from the EUP-modified pressure and energy density. As shown in Figure~10, $c_s^2(z)$ remains strictly positive and finite throughout the redshift range, confirming that the fluid does not exhibit any classical instabilities. The result implies that the EUP-induced effective fluid is dynamically and thermodynamically stable and capable of supporting causal perturbations.
	
	In summary, the EUP framework provides a theoretically motivated, quantum gravity inspired correction to cosmology that naturally gives rise to late-time acceleration and unifies gravitational dynamics with thermodynamic considerations. It offers an intriguing alternative to conventional dark energy models and merits further investigation, particularly in connection with observational constraints and structure formation.

\end{document}